Menoufiya University
Faculty of Computers & Information
Department of Information Systems

# Semantic Arabic Information Retrieval Framework

A Thesis Submitted in Partial Fulfillment of the Requirements for the
Degree of Master in Information Systems
To
Information Systems Department
Faculty of Computers and Information,
Menoufiya University

### By

## *Eissa Mohammed Mohsen Alshari*
Ibb University -Yemen

## *Supervisors*

### Dr. Hatem Mohammed Said Ahmed
*Chair of Information Systems Department
Faculty of Computers and Information,
Menoufiya University*

### Dr. Emad Elabd
*Information Systems Department
Faculty of Computers and Information,
Menoufiya University*

### (2014)

# Semantic Arabic Information Retrieval Framework

A Thesis Submitted in Partial Fulfillment of the Requirements for the
Degree of Master of Information Systems
To
Information Systems Department
Faculty of Computers and Information,
Menoufiya University

## By

*Eissa Mohammed Mohsen Alshari*
Ibb University -Yemen

## Referees and Judging Committee

| **Prof.Dr. Rafat Abdulftah Alkmar** | **Dr. Araby Alsaid Ibrahem** | **Dr. Hatem Mohammed Said** |
|---|---|---|
| *Professor of Computer Engineering, Faculty of Engineering, Shubra, Benha University* | *Dean of the Faculty of Computer and Information, Menoufiya University* | *Chair of Information Systems Department, Faculty of Computers and Information, Menoufiya University* |

**(2014)**

# Acknowledge

I am indebted to my advisor, Dr. Hatem Abdulkader and Dr.Emad Elabd, for the guidance they afforded me during my studies. They gave me plenty of freedom to explore the directions I was most interested in. At the same time, they constantly reminded me of the importance of focusing on the essential things and foregoing lower-hanging fruit that carry little benefit in the larger perspective.

Dr Emad's comments on the paper drafts were always to the point, and invariably led to much more coherent and simply better papers. From our countless discussions, I adopted a useful habit of always thinking about evaluation before launching a new study. He also supported me at the times when my spirit was low. He was always insisting that there is light at the end of the tunnel.

I am especially thankful to my wife for her unconditional love, for putting up with the countless hours I spent in my thesis work, and for being there when I needed here. I am deeply beholden to my dear father who cannot see these moments in my life. I am also thankful to my father for teaching me the value of knowledge and education. There is no words in any natural language could be sufficient to thank my great mother for all she had done for me. So this thesis is dedicated to them.

# Abstract

The continuous increasing in the amount of the published and stored information requires a special Information Retrieval (IR) frameworks to search and get information accurately and speedily.

Currently, keywords-based techniques are commonly used in information retrieval. However, a major drawback of the keywords approach is its inability of handling the polysemy and synonymy phenomenon of the natural language. For instance, the meanings of words and understanding of concepts differ in different communities. Same word use for different concepts (polysemy) or use different words for the same concept (synonymy).

Most of information retrieval frameworks have a weakness to deal with the semantics of the words in term of (indexing, Boolean model, Latent Semantic Analysis (LSA) , Latent semantic Index (LSI) and semantic ranking, etc.).

Traditional Arabic Information Retrieval (AIR) models performance insufficient with semantic queries, which deal with not only the keywords but also with the context of these keywords. Therefore, there is a need for a semantic information retrieval model with a semantic index structure and ranking algorithm based on semantic index.

In this Thesis, a Semantic Arabic Information Retrieval (SAIR) framework is proposed. This new framework merges between the traditional IR model and the semantic Web techniques. We have implemented the traditional AIR and SAIR frameworks. SAIR has semantic index differ index in traditional model. We add Reference Concept (RC) to traditional index. Thus, terms in semantic index has meanings more than traditional index. Then we have construct the ranking approach based on vector space methodology.

Finally, traditional model and semantic model performances are tested by measuring their precision, recall and run time. The obtained results from SAIR are compared with the results of the traditional IR model.

The simulation results of the proposed framework show a significant enhancement in terms of precision and recall but the run time is highly increased.

# Summary

Nowadays, the internet has large amounts of information and documents available in Arabic language. The most common commercial search engines such as Google and Yahoo support Arabic language. This support is mainly based on the classical Arabic language. In the other side, these search engines fail to get good results for Arabic query. The Arabic language is complex because of its complex syntax and the richness of its terms semantics. Our main objective in this research is to develop a framework for Arabic information retrieval based on the semantic. The proposed model based on a semantic data model for Arabic terms. Finally, we tested framework using a standard data set and checked the results using IR most known measurements such as the precision, recall and run time to evaluate our proposed model. The results of information retrieval with semantic Web enhanced when compared with traditional models.

# Contents

# List of Figures

# List of Abbreviation

| | |
|---|---|
| AIR | Arabic Information Retrieval |
| AVSM | Arabic VSM |
| AWN | Arabic WORDNET |
| D | Documents |
| DF | Document Frequency |
| *IDF* | Inverse Document Frequency |
| IR | Information Retrieval |
| IRI | Internationalized Resource Identifier |
| $I_t$ | Inverted List |
| KB | Knowledge Bases |
| LSA | Latent Semantic Analysis |
| LSI | Latent Semantic Index |
| MRR | Mean Reciprocal Rank |
| NLP | Natural Language Process |
| OWL | Web Ontology Language |
| Q | Query |
| QEQ | Query Expensive QE |
| RC | Reference Concept |
| RDF | Resource Description Framework |
| RDFS | Resource Description Framework Schema |
| RDQL | RDF Data Query Language |
| RST | Rhetorical Structure Theory |
| SAIR | Semantic Arabic IR |
| $SIM_C$ | Cosine Similarity Measure |
| SPARQL | SPARQL Protocol And RDF Query Language |
| SW | Semantic Web |
| T | Term |
| TF | Term Frequency |
| TF-IDF | Frequency–Inverse Document Frequency |
| URI | Uniformed Resource Identifier |
| VSM | Vector Space Model |
| WWW | World Wide Web |
| XML | Extensible Markup Language |

# List of Tables

# Chapter 1
# Introduction

# Chapter 1  Introduction

## 1.1  Introduction

Arabic language is one of the most widely spoken languages. It is one of the Semitic languages, that spoken by more than 422 million of people [1]. This language has a complex morphological structure and is considered as one of the most prolific languages in terms of article linguistic. Information retrieval is the process of finding all relevant documents responding to a query from mainly unstructured textual data. The science and practice of storing, searching and founding Arabic information within data is called Arabic Information Retrieval [2].

The area of Information Retrieval includes many studies that have been proposed to help users to retrieve information on their interests. The majority of the previously undertaken work describes methods and tools to process English language-based documents. The traditional model for information retrieval framework assumes that each document is represented by a set of keywords, so-called index terms. There are several features that distinguish the Arabic language from other languages. For example, the Arabic language is written from right to left, it has a complex morphological structure, Arabic is polysemous (i.e. the same word may have several meanings), and contains of a rich set of vocabulary[1].

Due to the complex morphology, polysemy, and the rich set of vocabulary of Arabic language, the traditional IR technologies do not work efficiently with Arabic collections[3]. Therefore, Semantic Web (SW) based IR technologies are nominated to overcome this problem in AIR.

Semantic Web will enable machines to comprehend semantic documents and data, not human speech and writings. It can assist the evolution of human knowledge as a whole. It draws conclusions about the Web page and improves the existing Web with machine-interpretable metadata that allows a computer program to understand what is a Web page. Information retrieval is an everyday problem that almost concerns everybody in our society. Therefore, information retrieval techniques can be improved by using semantic Web technologies [4].

## 1.2  Problem Definition

A main cause for this thesis is that currently consolidated content description and query processing techniques for Information Retrieval IR are based on keywords. So they provide limited capabilities to grasp and exploit the conceptualizations involved in user needs and content meanings [5], [6]. Arabic Language has some complex issues, which differ from the western languages:

- Written from right to left.
- It's different from Western languages especially at the morphological and spelling variations and the agglutination phenomenon [7].

To the best of our knowledge based on our survey, most of AIR frameworks have weakness to deal with semantics as the following:

- Due to the complex morphology, polysemy and the rich set of vocabulary, the IR technologies did not work efficiently with very large data collections [8].

- A big gap between the classic AIR approaches and the Semantic Web technologies [9].

- One of these problems is the lack of Arabic Boolean semantic in AIR model. Therefore, there is a trend to use semantic technologies to develop Boolean semantic Information Retrieval IR model.

- Arabic Information Retrieval models performance is insufficient with semantic queries, which deal with not only the keywords but also with the context of these keywords [10].

- Some researches attempted to bridge the gap between the AIR and the SW communities in the understanding and realization of semantic search [4], [11].

Therefore, there is a need for a semantic information retrieval model with a semantic index structure and ranking algorithm based on semantic index [12], [13].

## 1.3 Thesis Contributions

In this thesis, we Study of semantic search from the IR and SW fields, identifying fundamental limitations in the state of the art. Despite the large amount of work on conceptual search in the English IR field but a few in Arabic IR with semantic web. In this work, we discuss the strengths and weaknesses of the proposals towards the semantic search paradigm from both the Arabic IR and the Semantic Web fields. In addition, we present Arabic information retrieval with semantic framework. This thesis proposes the exploitation of ontologies to improve semantic retrieval in unstructured information. In addition:

i. Introducing a new design of Information retrieval based on semantic Web techniques.

ii. Introducing two models in IR based on Semantic Web: Boolean model and vector space model.

iii. Introducing some ontologies to extract the relation between the words and to extract the meaning among a phrase.

## 1.4 Thesis Structure

This thesis has been divided into three main parts. The first one in chapter 2&3 gives background for knowledge and a general literature survey of semantic search systems from both, the SW and IR areas. The second part in chapter 4 contains the design, implementation and evaluation of the semantic Arabic information retrieval model proposed in the thesis. The third part in chapter 5 contains experiment results. These main parts comprise several individual chapters, as follow:

**Chapter 2**: It provides a brief introduction of the Information Retrieval IR. This chapter provides a brief overview of the semantic-based knowledge technologies. It introduces the semantic knowledge concept as well as the advancements and problems on its representation, acquisition, annotation and evaluation.

**Chapter 3**: It provides a survey of the works that have attempted to solve the problem of semantic search in both, the IR and the SW areas.

**Chapter 4**: It presents our proposed semantic Arabic information retrieval model. The researchers provide a detailed description of how introducing a level of conceptualization in classical IR models can help to improve search over traditional keyword-based approaches.

**Chapter 5**: Validation of results in traditional model compared with proposed model.

**Chapter 6**: It discusses our conclusions and points out future research lines.

# Chapter 2

# Information Retrieval
# and
# Semantic Web

# Chapter 2  Information Retrieval
# and Semantic Web

This chapter provides a brief introduction of the IR and SW fields. The purpose of this chapter is to provide an overview focusing on the fundamental notions needed for later reference in the chapters where the thesis contribution is developed. Section 2.1 motivates the IR problem and discusses the complete IR process. Section 2.2 describes the semantic Web models.

## 2.1  Information Retrieval

The discipline that deals with retrieval of unstructured data is called Information Retrieval. IR is the process of finding all relevant documents responding to a query from mainly unstructured textual data [14]. Information Retrieval deals with relevant information items given specific information needs of users. As retrieval problems are defined in various environments such as the WWW, corporate knowledge bases or even personal desktops [15].

Information Retrieval focuses on retrieving documents based on the content of their unstructured components. An IR request (typically called a "query") may specify desired characteristics of both the structured and unstructured components of the documents to be retrieved, e.g., "The documents should be about 'Information retrieval' and their author must be 'Smith' ". In this example, the query asks for documents whose body (the unstructured part) is "about" a certain topic and whose author (a structured part) has a specified value [16].

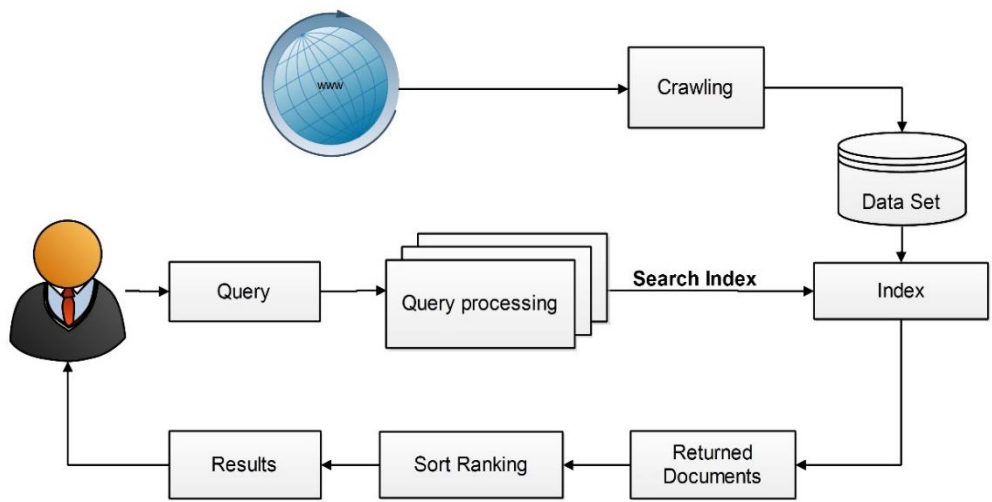

Figure 2-1 Information Retrieval model

Figure 2-1 shows the different components of the IR processes: Indexing, Query processing and Matching. More details about IR model will be explained later in this chapter [17].

### 2.1.1 Motivation of IR

Libraries have traditionally been the main information repositories of historical cultures. For example, the Ancient Library of Alexandria was founded around 280 BC by Ptolomeo I Soter with the purpose of preserving the Greek civilization, surrounded in Alexandria by a very conservative Egyptian civilization. It turned out to have around 700,000 scrolls.

Ptolomeo II commissioned the poet and philosopher Callimachus the task of cataloguing all books and volumes of the library. He was the first librarian of Alexandria and as a result of his work, Pinakes, the first thematic catalogue (to be known in our days) of history, was created. Other examples of big libraries are the Vatican Library that is created around 1500 B.C. and containing about 3,600 codices and the British Museum created around 1845 and containing about 240,000 books.

Nowadays, the amount of information available in document repositories has dramatically highly increased, and to a very large extent, it is stored in digital format. However, just because the content is available it does not mean that it is useful. Inversely, the user may not always find the information he may need. This problem arose already in the early days of computer technologies. In 1930 Vannevar Bush thought about a machine called Memex, "a device in which an individual stores all his books, records, and communications, and which is mechanized so that it may be consulted with exceeding speed and flexibility".

In 1950 Calvin Mooers coined the term Information Retrieval" but it was not until 1960, when Maron & Kuhns defined the problem of Information Retrieval as "adequately identifying the information content of documentary data". Following this idea, a lot of researches have been undertaken thereafter with the aim of making the information available in digital repositories universally accessible and effectively useful [9].

### 2.1.2 IR model processes

Information retrieval is one of the Natural Language Processing NLP applications. The goal of an IR system can be described as the representation, storage, organization of, and access to information items [2]. It has three main processes, namely [18], [19]: Indexing, Query processing and Matching (search and ranking) [17]. In indexing phase, documents are indexed using keywords that represent each document in the collection and extraction of item content features and descriptors into a logic representation of items [20]. In query reformulation stage, queries are reconfigured to comply with the model of information retrieval approach [2]. Finally, in the matching stage, the query inserted by user will be matched with index and the matched document are retrieved and ranked, based on its similarity with query. Figure 2-2 shows IR process, IR have two main process (Indexing, Search). In Index process, term will be extract from documents and store in inverted index [21]. In Search process, the user search a query and terms will extracts from this query. Thus, terms of query searches in inverted index. Finally the results of matching between inverted index and query will sort based on the ranking algorithms.

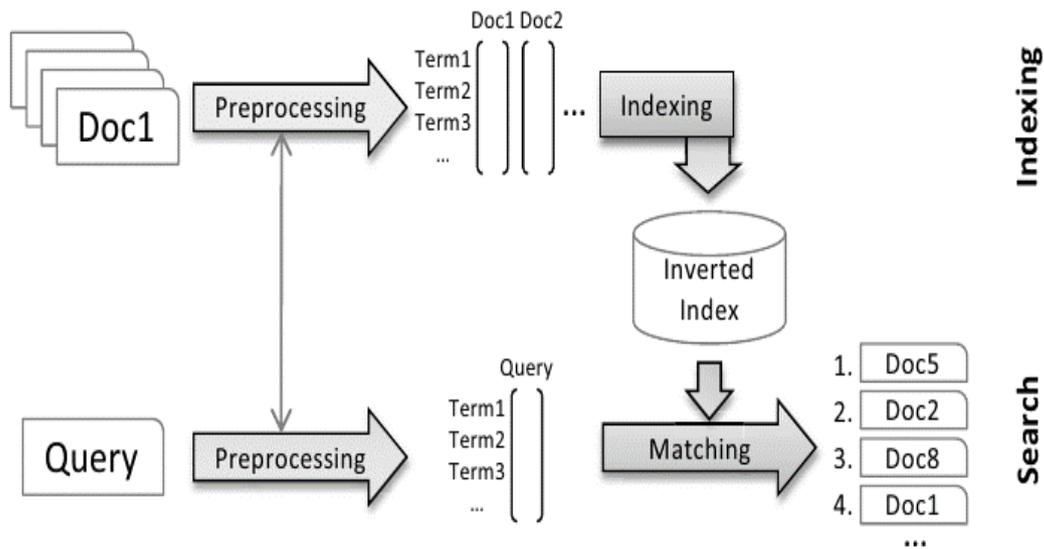

Figure 2-2 Information Retrieval process [17]

### 2.1.2.1 Indexing:

The information symbols extracted from collections by the analysis algorithms are stored and managed by the indexing module. Building an index from a document collection involves several steps, from gathering and identifying the actual documents to generating the final index [22].

The core element of an indexing mechanism is the inverted index that lists information symbols and all documents containing that symbol. The efficiency of indexes is affected by several design aspects: compression of the index reducing memory usage; tree structured indexes or hash-based indexes allowing a quicker look-up of the index table; sorting documents of an index entry limits the number of analyzed documents [23].

Not all the pieces of information item are equally significant for representing its meaning. In written language, for example, some words carry more meaning than others. Therefore, it is usually considered worthwhile to pre-process the information items to select the elements to be used as index objects. Indices are data structures that are constructed to speed up search. It is worthwhile building and maintaining an index when the item collection is large and semi-static. The most common indexing structure for text retrieval is the inverted file.

This structure is composed of two elements: the vocabulary and the term occurrences. The vocabulary is the set of all words in the text. For each word in the vocabulary a list of all the text positions where the word appears is stored. The set of all those lists is called occurrences [24].

## 2.1.2.2 Query processing:

The user needs, the query, are parsed and compiled into an internal form. In the case of textual retrieval, query terms are generally pre-processed by the same algorithms used to select the index objects. Additional query processing (e.g., query expansion, stop words and stemming) requires the use of external resources such as thesauri or taxonomies [25]. The most frequent words will most surely be the common words such as "the" or "and," which help build ideas but do not carry any significance themselves [26]. In fact, the several hundred most common words in Arabic and English (called stop words) are removed from query [27]. There is not one definite list of stop words, which all tools use and such a filter is not always used. Some tools specifically avoid removing them to support phrase search. On other hand, the stemming is one of query processing phase in IR model. Stemming use in a document indexes and queries. There are various approaches to stemming [28]. Stemming algorithms such as the Porter stemmer 1980 in English algorithms utilizes suffix stripping in a series of steps [29].

Lemmatizers identify the lexeme of a given word form, usually through dictionary lookup. N-gramming is another option requiring no linguistic knowledge or dictionaries and acts as a compound splitter and stemmer [30]. The basic stemming methods is [31]:

- Remove ending
    - If a word ends with a consonant other than s, followed by an s, then delete s.
    - If a word ends in es, drop the s.

- If a word ends in ing, delete the ing unless the remaining word consists only of one letter or of th.

- If a word ends with ed, preceded by a consonant, delete the ed unless this leaves only a single letter.

- …

- Transform words

- If a word ends with "ies" but not "eies" or "aies" then "ies --> y."

### 2.1.2.3 Matching:

User queries are matched against indexing terms. The result of this operation, a set of information items is returned to user [32].  Matching based on a set of roles between the user needs and information techniques. Thus, the set of information items returned by the matching constitutes an inexact. Therefore, the matching step need some algorithms to sort result, this step called ranking.

### 2.1.3 Information retrieval models

Information retrieval has three models, namely: Boolean model, vector space model and Probabilistic (Fuzzy) model. Each model determines documents representation in the index and thus controls the query reformulation and rank the matching results [9]. In Boolean model: a document either matches or does not match a query. It's a simple retrieval model based on set theory and Boolean algebra. Documents are represented by the index terms extracted from documents, and queries are Boolean expressions on terms. The vector space model (VSM) recognizes that the use of binary weights is too limiting and proposes a framework in which partial matching is possible. This is accomplished by assigning non-binary weights to index terms in queries and documents. These terms weights are ultimately used to compute the degree of similarity between each document stored in the system and the user query. By sorting the retrieved documents in decreasing order of this degree of similarity, the VSM takes into consideration documents which match the query terms only partially. The main resulting effect is that the ranked document answer set is considerably more precise (in the sense that it better matches the user information need) than the answer set retrieved by a Boolean model. The probabilistic model aims to capture the IR problem in a probabilistic framework. The fundamental idea is as follow. Given a query $q$ and a collection of documents $D$, a subset $R$ of $D$ is assumed to exist which contains exactly the relevant documents to $q$ (the ideal answer set). The probabilistic retrieval model then ranks documents in decreasing order of probability of belonging to this set (i.e. of being relevant to the information need), which is noted as P ($R$ /$q$, $dj$), where $dj$ is a document in $D$ [33].

### 2.1.3.1 Boolean model:

Boolean Models have been the first retrieval models used in the start of information retrieval which treats the user input query as an expression devised by Boolean logic. In the case of the Boolean retrieval model, relevance is binary and is computed by matching binary vectors representing term occurrence in the query to binary document vectors representing term occurrence [15]. The Boolean model algorithms in role (AND) is:

For each query term $t$

    Retrieve lexicon entry for $t$

    Note and address of $I_t$ (inverted list)

Sort query terms by increasing $f_t$

Initialize candidate set $C$ with $I_t$ of the term with the smallest $f_t$

For each remaining $t$

    Read $I_t$

    For each $d \in C$, if $d \notin I_t$, $C <- C - \{d\}$

    If $C = \{\}$, return… there are no relevant docs

Look up each $d \subset C$ and return to the user [24]

### 2.1.3.2 Vector Space model

Vector space models (VSM) are based on vector space representations of documents. Terms store in term-document matrix based on term frequencies. Functions computing scores for a single query term $t$ are based on the following measures:

- Term Frequency $tf$ in the document $tf_d(t)$.

- Document frequency $df$ of the query term $df(t)$.

- Number of documents in the collections $D$.

Currently, keywords-based techniques are commonly used in information retrieval. Among these keywords-based methods, Vector Space Models are the most widely adopted. Using VSM, a text document is represented by a vector of the frequencies of terms appearing in this document. The similarity between two text documents is measured as the cosine similarity between their term frequency vectors; however, a major drawback of the

keywords-based VSM approach is its inability of handling the polysemy and synonymy phenomenon of the natural language [12]. As meanings of words and understanding of concepts differ in different communities, different users might use the same word for different concepts (polysemy) or use different words for the same concept (synonymy). We will discuss more details about VSM in chapter 4.

### 2.1.3.3 Fuzzy model

Although a model of probabilistic indexing was proposed and tested by Maron and Kuhns (1960), the major probabilistic model in use today was developed by Robertson and Sparck Jones (1976) [34]. This model is based on the premise terms that appear in previously retrieved relevant documents for a given query that should be given a higher weight than if they had not appeared in those relevant documents. In particular, they presented the following table showing the distribution of term $t$ in relevant and non-relevant documents for query $q$.

|  | | Document Relevance | |
|---|---|---|---|
|  | $+$ | $-$ | |
| $+$ | $r$ | $n - r$ | $n$ |
| $-$ | $R - r$ | $N - n - R + r$ | $N - n$ |
|  | $R$ | $N - R$ | $N$ |

(Document Indexing on the left rows)

Figure 2-3 Index in fuzzy model

$N$ = the number of documents in the collection

$R$ = the number of relevant documents for query $q$

$n$ = the number of documents having term $t$

$r$ = the number of relevant documents having term $t$

They then use this table to derive four formulas that reflect the relative distribution of terms in the relevant and non-relevant documents, and propose that these formulas are used for term-weighting (the logs are related to actual use of the formulas in term-weighting) [35].

### 2.1.4 Evaluation of IR model

The final goal of IR evaluation is satisfaction human about retrieved documents. Relevance is an inherently subjective concept [36] in the sense that satisfaction of human needs is the ultimate goal, and hence the judgment of human users as to how well retrieved documents satisfy their needs is the ultimate criterion of relevance. Therefore, human beings often disagree about whether a given document is relevant to a given query. In general, disagreement among human judges is even more likely when the question is not absolute relevance but degree of relevance. Therefore, the relevance to the user's query, differs about pertinence to the user's needs [37],[38]. IR has two success measures, both based on the concept of relevance (to a given query or information need), are widely used: "precision" and "recall". Precision is defined as, "the ratio of relevant items retrieved to all items retrieved, or the probability given that an item is retrieved that it will be relevant" [38]. Recall is defined as, "the ratio of relevant items retrieved to all relevant items in a file (i.e., collection), or the probability given that an item is relevant that it will be retrieved" [37]. Other measures have been proposed, but these are by far the most widely used. Recall is more difficult than precision because it depends on knowing the number of relevant documents in the entire collection, which means that all the documents in the entire collection must be assessed, If the collection is large, this is not feasible. However, Precision need set of competent users or judges agree on the relevance or non-relevance of each of the retrieved documents.

## 2.2 Semantic Web

Matching only keywords may not accurately reveal the semantic similarity among text documents or between search criteria and text documents. Due to the heterogeneity and independency of data sources and data repositories. For example, the keyword "java" can represent three different concepts: coffee, an island, or a programming language, while keywords "dog" and "canine" may represent the same concept in different documents[39].

Semantic Web will enable machines to comprehend semantic documents and data, not human speech and writings. It can assist the evolution of human knowledge as a whole[14]. As a technology, the Semantic Web can be summarized as "knowledge representation meets the Web" [43]. The goal is to create declarative representational notations, i.e. languages, that would enable automatic processing and composition of information in the Web.

The world wide Web (WWW) has changed the way of communication among the people and the way of conducting businesses. The present Web's contents represent the information to be more human readable and understandable rather than machine readable. The semantic Web is the Web of data rather than the Web of documents. Semantic Web is machine readable [34]. Adding semantics to Web site structure makes the Web site code readable by both humans and machines.

The semantic Web contains meta-data, which is data about data and it contains ontologies. Ontology is an agreement needed to be added to the Web page to let the machine understands the document [16].

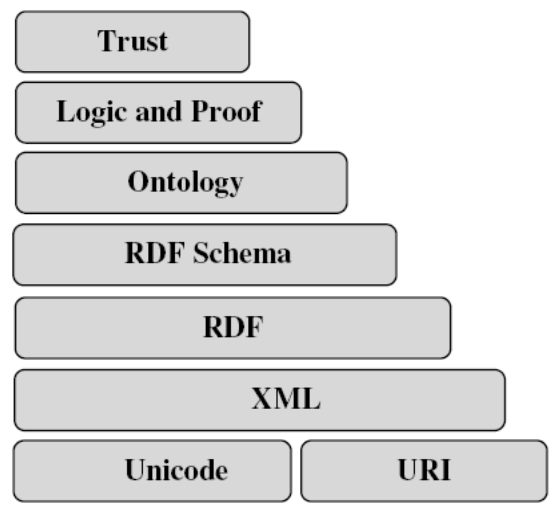

Figure 2-4 The layer cake of semantic Web

Figure 2-4 describe the semantic Web layers .The layered model for semantic technologies contains an illustration of the hierarchy of semantic stack, where each layer exploits and uses capabilities of the layers below:

- Internationalized Resource Identifier (IRI), generalization of Uniformed Resource Identifier (URI), provides means for uniquely identifying ontological resources.

- Extensible Markup Language (XML) is a markup language that enables creation documents of structured data.

- XML Namespaces provides a way to use markups from different

sources. They are used to refer to different sources in one document.

- Resource Description Framework (RDF) is a language for creating a data model for objects (or "resources") and relations among them. It enables to represent information in the form of graph.

- Resource Description Framework Schema (RDFS) provides basic vocabulary for describing properties and classes of RDF resources. Using RDFS it is possible to create hierarchies of classes and properties.

- Web Ontology Language (OWL) extends RDFS by adding more advanced constructs to describe the semantics of RDF statements.

- RDF Data Query Language (RDQL) and SPARQL Protocol and RDF Query Language (SPARQL) are ontology query languages. They used to extract specific information from RDF graphs.

- Cryptography is important to ensure and verify that RDF statements are coming from trusted sources. This can be achieved by appropriate digital signature of RDF statements.

- Trust to derived statements will be supported by (a) verifying that the premises come from trusted sources and by (b) relying on a formal logic for deriving new information.

- User interface is the final layer that will enable humans to use ontology-based semantic applications and therefore to exploit ontology-based semantic knowledge.

The proposed model in chapter 4 depends on RDF, RDFs, and ontology mainly; therefore, we will focus about this technique extensively.

### 2.2.1 **RDF and RDFs**

RDF has a very simple data model and gives users the opportunity to describe the resources by their OWL ontology by using the RDFS language. RDFS is responsible for defining the vocabulary of domain [11]. By using metadata and ontologies, semantic technology adds meaning to the Web page [16]. Figure 2-5 shows the difference between RDF and RDFs. The RDF Schema RDFS enriches the data model, adding vocabulary and associated semantics for Classes, subclasses, Properties and sub-properties.

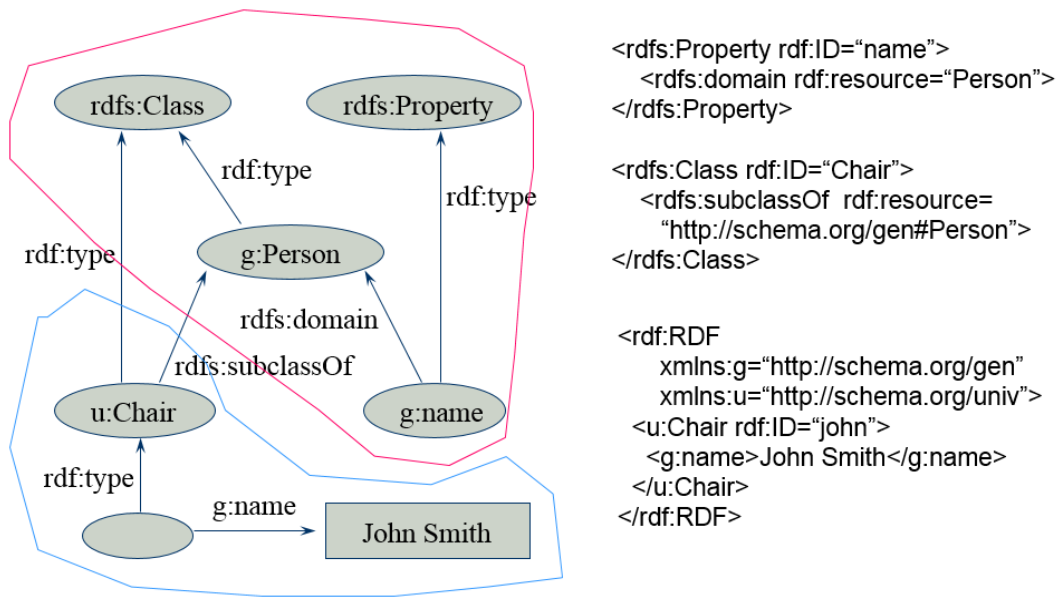

Figure 2-5 RDF and RDF schema

### 2.2.2 **Ontology**

Ontology is one of the most important knowledge representation techniques in semantic Web. Kumar defines ontology as a knowledge that provides semantic for understanding the meaning of data [40]. Ontology is an explicit specification of a representational vocabulary for a shared domain of discourse definitions of classes, relations, functions, constraints and objects [41]. The main purpose of building domain ontology is to mimic how the human brain keeps the semantics stored [4].

The OWL is a well-known class of ontology [42]. The term "ontology" originates from philosophy as "the study of the nature of existence" [11], which is about describing the things that exist in the world around us. In computer science, ontology has a different definition: "an explicit and formal specification of a conceptualization" [35].

# Chapter 3
# Related Work

# Chapter 3  Related Work

This chapter provides a brief related work about information retrieval and semantic fields. The indexing presents and explains in (Section 3.1) and the matching (Section 3.2). The IR and SW are discussed in (Section 3.3). Finally, the ontology is presented in Section 3.3.

## 3.1  Indexing

Indexing is one of information retrieval phases. It has many researches, which it discusses the storing and constructing. Ataa Allah, et al. [1] studied the syntagmatic knowledge impact on the latent semantic analysis for the information retrieval in a specialized Arabic corpus. They tried to improve Arabic Information Retrieval AIR by using noun phrases in the indexation process. Nevertheless, that did not show any improvement of the IR system performance.

Al-Jedady et al. [43] presented a technique to encode index terms using 6-bits length coding which gives 64 different possibilities and 33 codes for encoding the 28 Arabic characters + 5 different variations of some characters. The spilt and encode term is called Bigram index term coding . The indexer builds one or more index files to speed up the searching process. Encoded index-using bigram coding scaling-up by 50% of queries using the same resources, without investing in new resources. The presented Index term compression show a significant reduction on the number of comparisons needed for binary search. Their proposed technique also showed a good reduction of terms' size, which contributes in the reduction of the overall index size. It also showed a good reduction of the number of comparisons needed for sequential search.

Mansour et al.[13] proposed a method based on morphological analysis and on a technique for assigning weights to words. They addressed the information retrieval problem of auto-indexing Arabic documents. Auto indexing a text document refers to automatically extracting words that are suitable for building an index for the document. The morphological analysis uses a number of grammatical rules to extract stem words that become candidate index words. There are two types of indexing: thesaurus based indexing and full-text based indexing.

Ibrahim et al. [44] presented a framework for the application of Rhetorical structure theory (RST) in the Arabic language, in order to improve the ability to extract meanings behind the text. RST is a descriptive theory of a major aspect of the structure of natural text. Average Precision 34%, which is better than other commercial systems that show mean Average Precision 13%.

## 3.2 Matching

Search about ambiguous words in IR approaches are complex because their diversity and large number of dimensions involved in the information search task. ANIS et al. [6] proposed a new approach for determining the adequate sense of Arabic words. The proposal extract the contexts from corpus, they applied measures of similarities in information retrieval methods (Okapi[45], Harman[46], and Croft[47]) to allow the system to choose the context using the most closer to the current context of the ambiguous word. They applied Lesk algorithm to distinguish the exact sense of the different senses given by measures of similarity [48]. The result of each comparison is a score indicating the degree of semantic similarity between a sentence (containing an ambiguous word) and a document (that represents the contexts of use for a given sense of the ambiguous word). They used Lesk algorithm as a measure method and the obtained accuracy rate was only 73% [48]. It used some inefficient algorithms such as *"New approach for extracting Arabic roots"*, Al-Shalabi-Kanaan [49] to extract the stems of the Arabic words, which achieved only 14% when Al-Shawakfa [50] compared some Arabic root finding algorithms. Therefore, their approach achieves precision of 78% and recall of 65% only.

Froud et al. [51] used the well-known abstractive model - Latent Semantic Analysis LSA - with a wide variety of distance functions and similarity measures to measure the similarity between Arabic words, such as the Euclidean Distance, Cosine Similarity, Jaccard Coefficient, and the Pearson Correlation Coefficient. They used LSA with and without stemming in two different data set to know how stemming impact on the meaning. They show that the use of negatively stemming affects the obtained results with LSA model, when it tries to measure the similarity between two different words that have the same root.

Paralic et al. [52], compared between traditional full text search based on vector IR model and the Latent Semantic Indexing method that use ontology-based retrieval mechanism. They developed package with three different approaches to document retrieval: vector representation, latent semantic indexing method LSI, and ontology-based method that is used in the Webocrat system. The approach describes the Webocrat-like approach that uses ontology for document retrieval purposes. Their experiments showed that the Webocrat-like approach based on an ontology is very promising, providing better retrieval efficiency than LSI or standard full text approach. However, as mentioned above, manual assignment of concepts to query has been used. They did not consider the relation in ontology for calculation of similarity between concepts. Moreover, they assumed that the set of relevant concepts to a query is known. On the contrary, the type of relation and set of relevant concepts are un-known to the model untested. The semantic index proposed requires implementing and evaluating. In addition, it addresses outline in technique only.

## 3.3  IR and SW

Although Information retrieval technology has been central to the success of the Web. Information Retrieval need many of researches to deal with meaning and concepts. Therefore, information retrieval and semantic Web requires fill the gap between IR and SW.

Fernández et al. [9] attempted to bridge the gap between the IR and the SW communities in the understanding and realization of semantic search. They proposed the generation of a novel semantic search model that integrates and exploits highly formalized semantic knowledge in the form of ontologies and Knowledge Bases KBs within traditional IR ranking models.

Table 3-1 summarized the most known approaches that integrate the semantic Web technologies with IR and their limitations. From the table, there is a big gap between the classic IR approaches and the Semantic Web technologies. One of these problems is the lack of Boolean semantic IR model. Therefore, there is a trend to use semantic technologies to develop

Boolean semantic IR model. Besides, the listed approaches show lack of standard evaluation semantic frameworks, semantic ranking, and multimedia based ontology.

Table 3-1. Limitations of semantic search approaches

| Criterion | Approaches | Limitation | IR | Semantic |
|---|---|---|---|---|
| *Semantic knowledge representation* | Statistical Linguistic conceptualization Ontology-based | No exploitation of the full potential of an ontological language, beyond those that could be reduced to conventional classification schemes | ☒ | partially |
| *Scope* | Web search Limited domain repositories Desktop search | No scalability to large and heterogeneous repositories of documents | ✓ | ☒ |
| *Goal* | - | Boolean retrieval models where the information retrieval problem is reduced to a data retrieval task | ✓ | ☒ |
| *Query* | Keyword query Natural language query Controlled natural language query Structured query based on ontology query languages | Limited usability | ✓ | ☒ |

| Criterion | Approaches | Limitation | IR | Semantic |
|---|---|---|---|---|
| *Content retrieved* | Data retrieval Information retrieval | Focus on textual content: no management of different formats (multimedia) | Partially | partially |
| *Content ranking* | No ranking Keyword-based ranking Semantic-based ranking | Lack of semantic ranking criterion. The ranking (if provided) relies on keyword-based approaches | ⊠ | ⊠ |
| *Coverage* | - | Knowledge incompleteness | Partially | ⊠ |
| *Evaluation* | - | Lack of standard evaluation frameworks | ✓ | ⊠ |

✓ exists ⊠ not exists

El-Shishtawy et al. [53] presented an Arabic summarization algorithm for extracting relevant sentences from free texts. The system exploits statistical and linguistic features to identify important keyphrases. keyphrases are automatically extracted from a document text are used to evaluate the importance of each sentence in the document.

Although there are numerous techniques for sentence level extraction, little attention is paid to the changing extraction strategy to achieve one or more summarization goals. In general, there are two methods for automatic text summarization: extractive and abstractive.

The algorithm addressed the extractive summarization involves copying significant units (usually sentences) of the original documents. However, abstraction summary is to produce summaries that read as text by humans. Therefore, abstraction summary needs the building of a semantic representation, the use of natural language generation techniques, the compression of sentences, the reformulation, or the use of new word sequences that are not presented in the original document. These methods are need semantic technology to deal with. The RDF, RDFs, and ontology can be used in the abstractive method.

Abouenour et al. [3] proposed semantic Query Expensive QE (QEQ) based on Arabic WordNet (AWN). The proposal has two types of experiment conducted: the keyword-based evaluation which uses a classical search engine as passage retrieval system, and the structure-based evaluation that uses the Java Information Retrieval System JIRS. It aimed at confirm the preliminary experiments which showed that the accuracy and the Mean Reciprocal Rank (MRR) have been improved and that semantic QE process (based on the current release of AWN) is adequate to improve the passage retrieval stage of an Arabic Q/A system. Also the semantic QE approach improves both the accuracy and the MRR. In addition, in the case where it is combined with JIRS. The approach has obtained an accuracy around 19.51% and 7,85% as MRR. Probability of relevant passage improved because they take into account the semantic and the structure of the question. In addition, the AWN project did not cover totally the standard Arabic version of AWN. It included WordNet only in ontology, because the other Arabic ontology techniques such as domain based ontology is difficult to measure.

### 3.4 Web Ontology Language

Web Ontology Language OWL is a family of knowledge representation languages or ontology languages for authoring ontologies or knowledge bases. The languages are characterised by formal semantics and RDF/XML-based serializations for the Semantic Web. In this chapter, we address many of research about construct and ontology approaches.

Hoseini. [54] described a Derivational Arabic Ontology used to model the Arabic Language. The knowledge is then retrieved when needed for using in computer-based applications mainly. The key idea underlying compositional approach is that the meaning of a sentence can be composed from the meaning of its syntactical constituents. In this work, the semantic representation of Arabic syntactical phrase is function of its constituent words and phrases. The automatic ontology constructions use the list of existing Arabic verbs to generate all. Its derivations populate the ontology in an easy and straightforward manner.

The proposal can be used as the perfect Arabic morphology analyzer. Strong morphology system will help the development of many applications such as information retrieval. This model needs a lot of study and application to assess the efficiency and performance. It did not specify ontology and semantic techniques that can be employed.

Al-Rajebah et al. [55] presented a new approach to build ontological models for Arabic language. They proposed their ontological model to be applied on Arabic Wikipedia to extract for each article its semantic relations using its info box and list of categories. The approach relies upon the semantic field theory such that any Wikipedia article is analyzed to extract semantic relations using its info box and the list of categories.

The approach evaluated using insufficient measures: human judges and precision whilst organizes ontology evaluation methods requires two dimensions: ontology quality criteria (accuracy, adaptability, clarity, completeness, computational efficiency, conciseness, consistency, and organizational fitness) and ontology aspects (vocabulary, syntax, structure, semantics, representation, and context)[56].

Zaidi et al. [7] described a Web-based multilingual tool for Arabic information retrieval based on ontology in the legal domain. It illustrated manual construction of the ontology and the way it edited using Protégé2000. Using Arabic documents identify the legal terms and the semantic relations between them before mapping them onto their position in the ontology. The attempted approach is made to improve the precision of the search thus mini missing the level of noise in the results. A set of query words is used to enable the machine translation of the query from Arabic into English and from Arabic into French.

Mazari et al. [57] proposed an approach of automatic construction that uses statistical techniques to extract elements of ontology from Arabic texts by reused information extraction techniques for extracting new terms that will denote elements of the ontology (concept, relation). To analyze the texts of the corpus, two statistical methods were used, the "repeated segments" to identify the candidate terms and "co-occurrence" to the updating of ontology. They formed a domain corpus by the recovery of text from articles of journals and books of the domain and also the collection of documents over the Web.

Beseiso et al. [58] evaluated the support of some tools such as Protégé and Jena, Sesame, and KOAN for Arabic language .

As shown in table 3-2, Arabic information retrieval and semantic is not supported by KAON2. Protégé and Sesame limited support. However, Jena is support better support Arabic language. Therefore, the current tools are not sufficient and many IR phases such as indexing, querying and crawling are not evaluated.

Table 3-2. Arabic tools support

| Tool | RDF | OWL | Query |
|---|---|---|---|
| Protégé | Support | Limited Support | Limited Support |
| Jena | Support | Support | Limited Support |
| Sesame | Limited Support | Limited Support | NO Support |
| KAON2 | NO Support | NO Support | NO Support |

The AIR need new tools to be developed to support the Arabic language natural language process NLP are critical. Moreover, development and design of semantic tools that supported Arabic language processing & encoding.

Aliane et al. [59] presented a project to build an ontology centered infrastructure for Arabic resources and applications. It aims at reusing ontology for creating tools and resources for both linguists and NLP researchers. They used Python language for implementing the extraction system. They opted for a statistical approach, namely the method of repeated segments calculation combined with some prior processing of the texts that comprise: segmentation, light stemming, stop words elimination. Al-khalil is OWL ontology under development. They baptized the project Al-Khalil in the sake of the famous grammarian AL-Khalil Ibn Ahmad Alfarahidi.

Khalifa et al. [60] presented project for building a framework for recognizing and identifying Arabic semantic opposition terms using Natural Language processing armed with domain ontologies. Semantic opposition is based on the concept of semantic fields/domains. They classified the Holy Quran into speech recognition, stop words, morphology analysis and ontology engine.

The framework requires evaluated usefulness and effectiveness via the judgment of human experts and through comparing it with more traditional approaches i.e. dictionaries. SemQ is a framework that is taken as an input a Quranic verse (i.e. sentence) and outputs the list of semantically opposed words in the verse along with their degree of opposition.

Aliane, H [8] presented an ontology based approach for multilingual information retrieval that has been implemented for Arabic, French and English. They proposed system based on knowledge representation formalism, namely semantic graphs, which support domain ontology. The domain ontology constitutes the kernel of the system and is used for both indexing and retrieval.

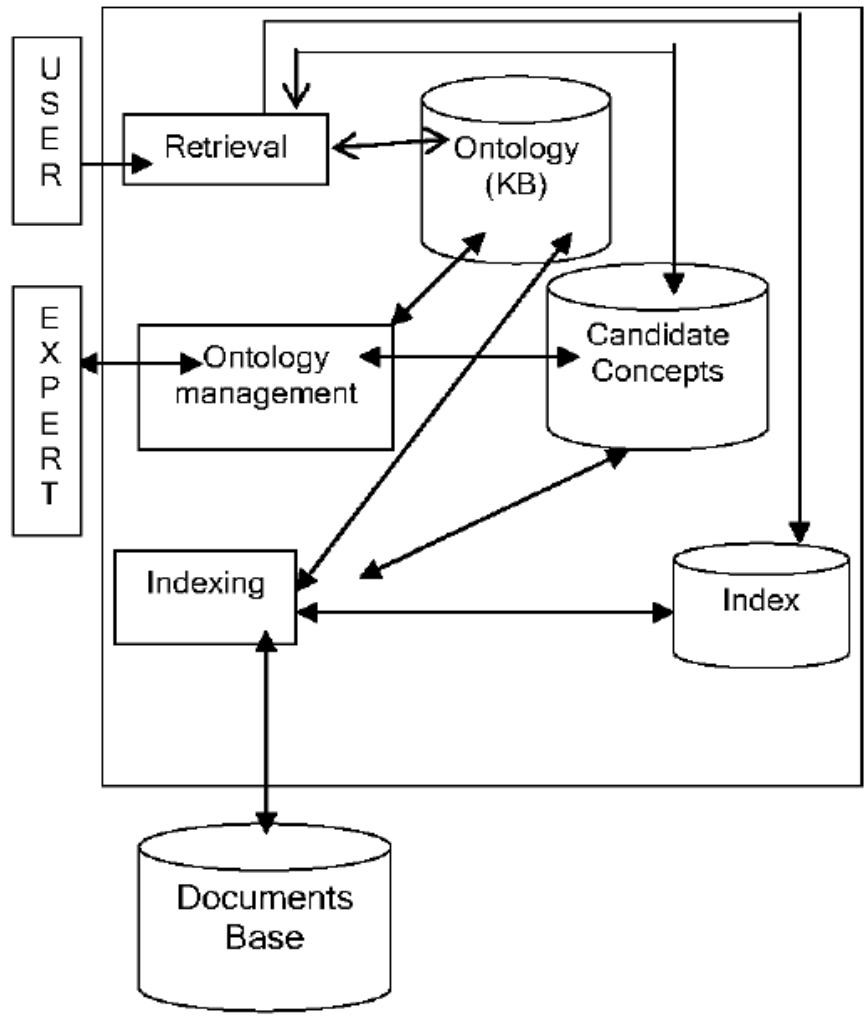

Figure 3-1 Architecture of the system [8]

The system has been developed using JAVA language in order to run on both windows and Unix platforms and documents are represented in XML format. Two kinds of interfaces are offered for the expert user who create, manage and update the ontology and for the end user who searches for documents. The interfaces are trilingual. The user can work with the language of his choice Arabic, French or English. The difficult task for ordinary people who are not familiar with the ontology however, the expert people in Arabic is insufficient.

# Chapter 4
# Proposed Model

# Chapter 4  Proposed Model

This chapter presents and explains the proposed approach and provides the details about the Boolean semantic model (Section 4.1) and the vector space model (Section 4.2). The indexing and semantic query processing phases are discussed in Section 4.1.1 and Section 4.1.2 respectively. Finally, the ontology construction process are discussed in Section 4.1.3.

## 4.1  Boolean Semantic IR

The key idea is to build a semantic inverted index to store not only words but also Reference Concepts (RC) reflex the meaning of these words in there phrases context. The reference ontology concept of a word is determined by getting a major concept links between all the words in the phrase. Therefore, it is based on all the terms of the phrase. In other words, all the words in the same phrase have the same reference ontology concept. The proposed model consists of two main parts: semantic  inverted index construction and semantic  query processing and retrieval.

### 4.1.1 Indexing phase:

In this phase, the semantic inverted index of a collection of documents is built. The algorithm of the index creation starts to manipulate each document of the collection by extracting and preprocessing its phrases one after another. The preprocessing operations on the phrase include the removal of the stop words which are listed in the stop words list and the stemming. These preprocessing operations are standard operations in any information retrieval system. The next operation is the reasoning of the ontology using the set of words that are resulted after the phrase preprocessing operation to get a reference concept from the ontology links between these words. Finally, each word of the phrase is stored in the semantic inverted index in the form [word, reference concept, DocID] where the DocID is a unique identifier for the document that this phrase and this word are belongs to. Algorithm 1 shows the pseudo code for performing this indexing process. The proposed model with an example is shown in Figure 4-1.

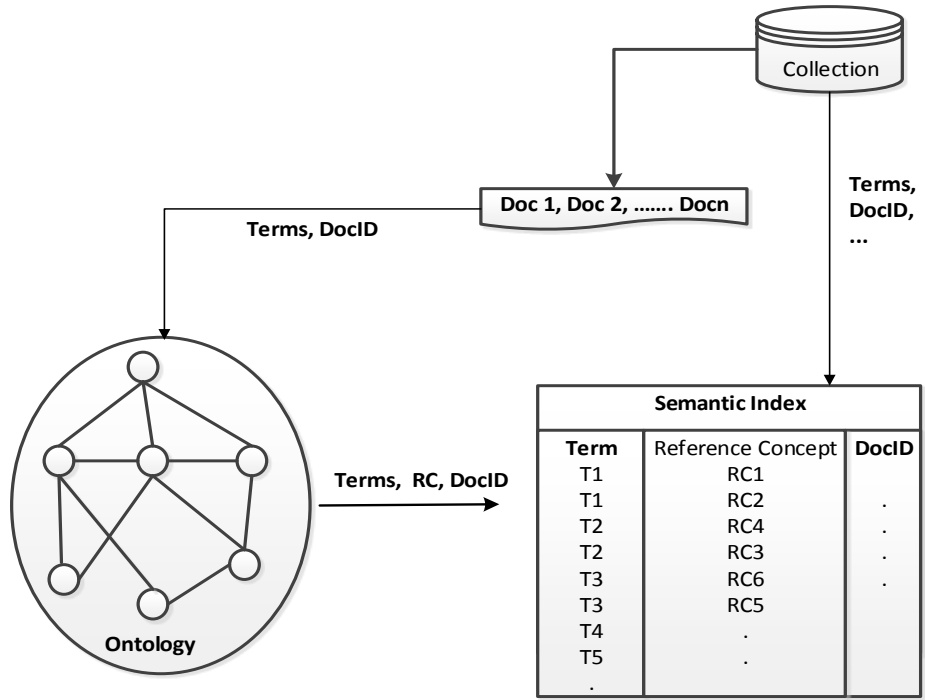

Figure 4-1 Semantic Index

### 4.1.2 **Semantic query processing and retrieval**

In this phase, the user's query is processed and the semantic inverted index is used to retrieve the required documents. The query can be a word or a phrase consists of a set of words. In case of only one word, the only preprocessing operation is the stemming and then the information retrieval engine searches in the inverted index for that word and returns to the user the set of documents that contains this word. In this case, if the word is stored in the semantic index with different reference concepts then the returned documents are organized based on the reference concepts to enable the user to select results based on his needs (i.e., in which context he wants his results?). In case of phrase query, this query is preprocessed by removing the stop words and stemming each word and then check the same ontology, which is used in the indexing phase using the set of words of the query phrase, and get the reference concept for these words.

The previous operation is the same operations that are applied to each phrase on the documents of the collection in the indexing phase. The result of this operation is a set of terms (words) and each term has his reference concepts, which is the same for all the terms of the query phrase. The next step is to match the terms of the query with the terms of the semantic index. The returned terms will be attached with their RCs. These results are filtered using the ontology by returning only the terms with RCs that have a relation with the RCs of query terms. Finally, the filtered results are returned to the user. Algorithm 2 shows the pseudo code for performing this query processing and retrieval process. Figure 4-2 shows an example of this process where the semantic query reference concept is RC and the equivalent terms have RC1, RC3, and RC6. The filter operation tries to decide if there is a relation between RC and (RC1, RC2, RC3, RC4, RC5, RC6).

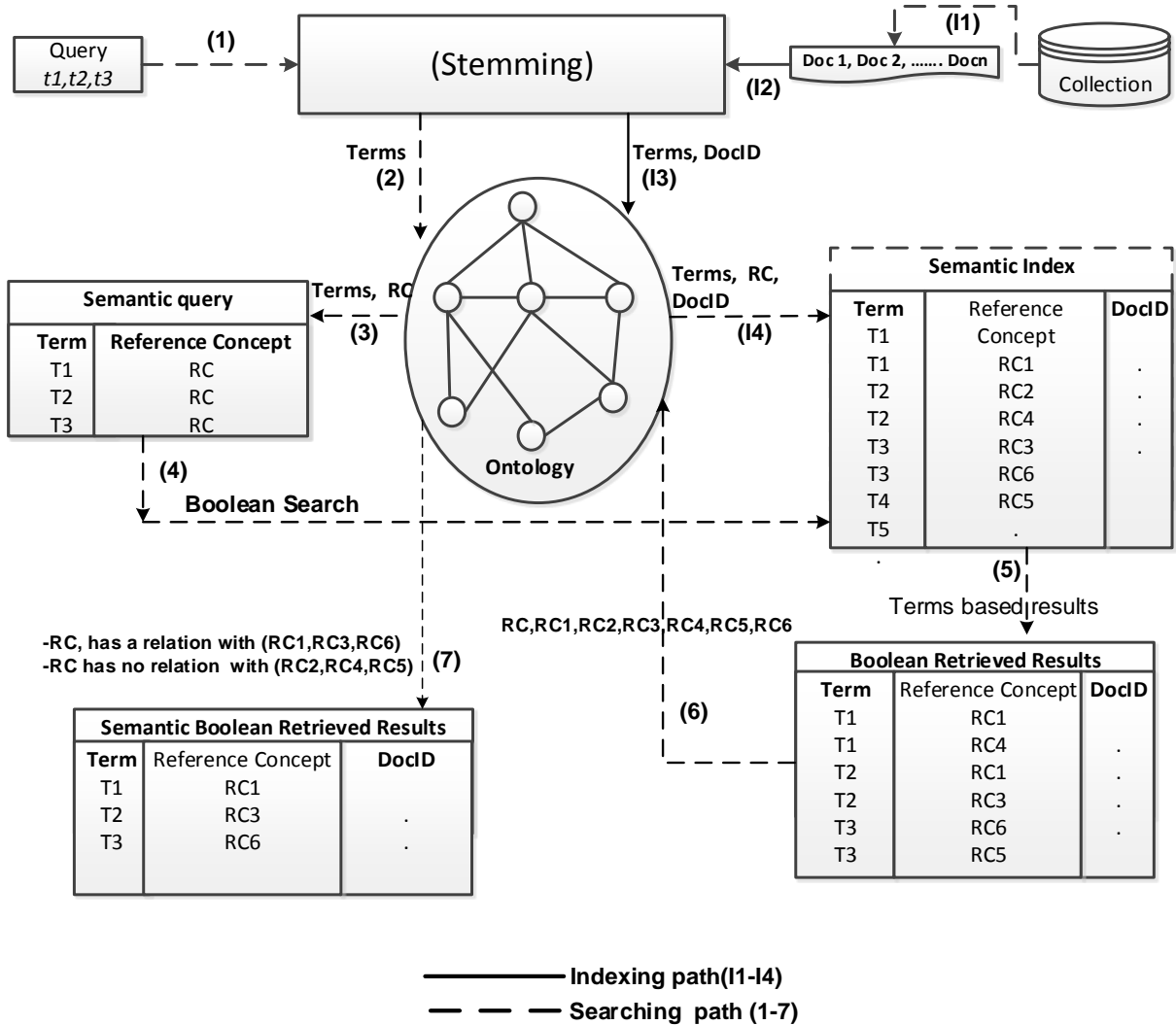

Figure 4-2 proposed approach

## Algorithm 1 Semantic inverted index -Indexing phase (a collection of documents and ontology)

#Let CDoc represents the collection of the documents {$Doc_1,…, Doc_n$}, Where $Doc_i \in CDoc$ and n is the number of the documents in the collection.

#Let $Doc_i$ represents a document that consists of a set of phrases {$Phr_{i1},…, Phr_{im}$} Where $Phr_{ij} \in Doc_i$ and m is the number of phrases in document

*$Doc_i$.*

*#Let $Phr_{ij}$ represents a phrase that consists of a set of words $\{w_{ij1}, ..., w_{ijl}\}$*

*Where $w_{ijk} \in Phr_{ij}$ and $l$ is the number of words in phrase $Phr_{ij}$.*

*#Let $Ont$ represents the used ontology and $RC_{ij}$ represents the reference*

*concept for the words of the phrase $Phr_{ij}$.*

*#Let $DocID_i$ is the DocID of document $Doc_i$.*

*For each $Doc_i \in CDoc$*

*{*

*For each $Phr_{ij} \in Doc_i$*

*{*

*Remove stop list*

*Stemming each $w_{ijk} \in Phr_{ij}$*

*Reasoning the ontology $Ont$ by the words $w_{ijk} \in Phr_{ij}$ and get the $RC_{ij}$*

*For each $w_{ijk} \in Phr_{ij}$*

*{*

*Store $[w_{ijk}, RC_{ij}, DocID_i]$ in the semantic inverted index*

*}*

*}*

*}*

*Return (semantic inverted index)*

**Algorithm 2. Query processing and retrieving (semantic inverted index, ontology, and user phrase).**

*#Let $QPh$ represents a query phrase that consists of a set of words $\{w_1, ..., w_l\}$ Where $w_k \in QPh$ and $l$ is the number of words in query phrase $QPh$.*

*#Let $Ont$ represents the used ontology and $RC_{ij}$ represents the reference concept for the words of the phrase $QPh$.*

*#Let $DocID_i$ is the DocID of document $Doc_i$.*

*Read query phrase QPh.*

*Remove stop list.*

*Stemming each $w_k \in QPh$.*

*Reasoning ontology Ont by word $w_k \in QPh$ and get the RC.*

*For each $w_k \in QPh$.*

*{*

  *Get the [$w_k$, $RC_i$, DocID] from the semantic inverted index.*

  *Reasoning the ontology Ont by $RC_i$ and RC.*

  *If there is a relation between $RC_i$ and RC then.*

   *{*

   *Retrieve [$w_k$, $RC_i$, DocID] to the user.*

   *}*

 *}*

    *Return (List of query words with its corresponding DocID).*

### 4.1.3 Ontologies construction

In this phrase, we suggested some ontologies to achieve, implement, and test our model. First, we create five ontologies. Arabic language has three ontologies (علوم - إلكترونيات  -  طبيعة ). English language has tow ontologies (Device and Natural).  The ontologies have some classes, properties and relation between it. We created ontologies in protégé tools. Protégé is a free, open source ontology editor and knowledge-base framework.

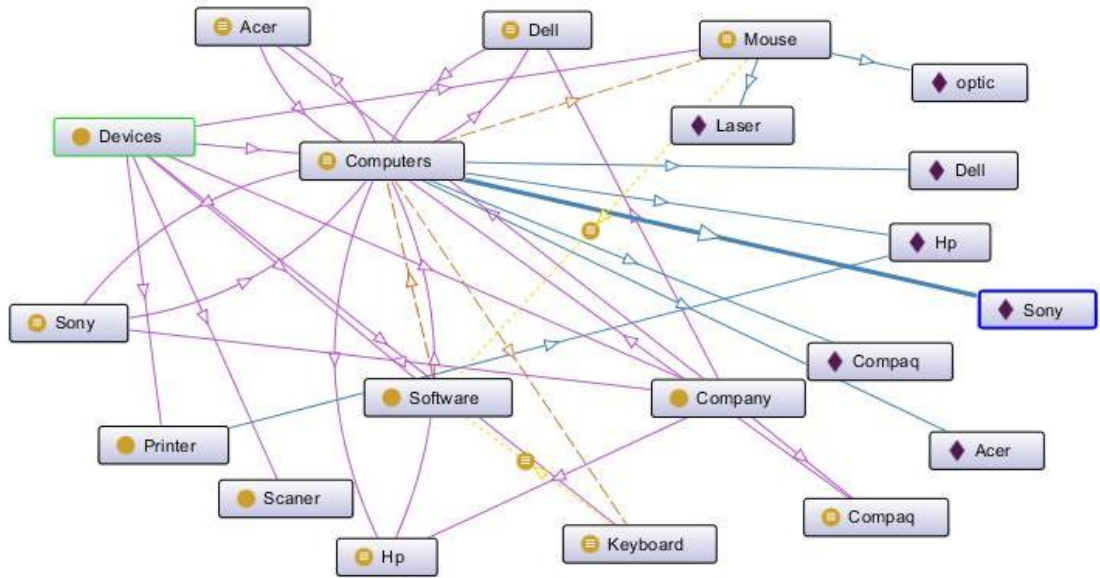

**Figure 4-3** Catch from English Ontology

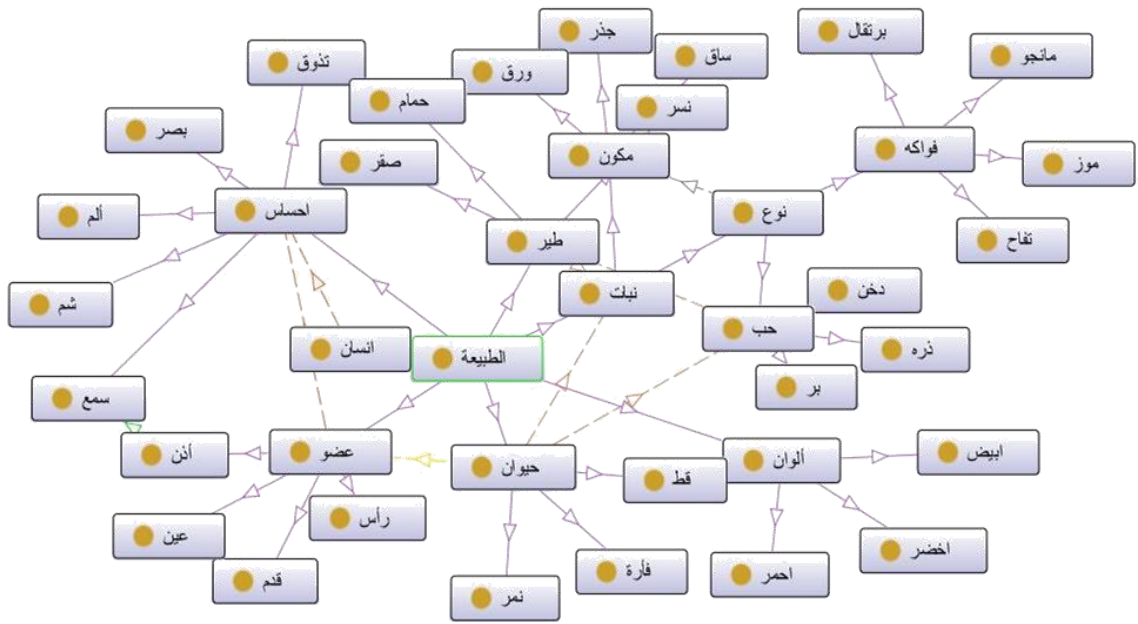

Figure 4-4 Catch from Arabic Ontology

Figure 4-3 and 4-4 shows part of English and Arabic ontologies. Then we need to convert ontologies form knowledge base in protégé to RDFs.

Because on RDFs used as a general method for conceptual description or modeling of information that implemented in this model using java language. Figure 4-5 shows RDFs in Jave language relies Jena tools, because protégé limited supported in Java.

```
<!-- http://www.semanticweb.org/maysaa/ontologies/2012/11/untitled-ontology-7# -->
<owl:ObjectProperty
      rdf:about="http://www.semanticweb.org/maysaa/ontologies/2012/11/untitled ontology-7# ">
<rdfs:domain
      rdf:resource="http://www.semanticweb.org/maysaa/ontologies/2012/11/untitled-ontology-7#أذن"/>
<rdfs:range
      rdf:resource="http://www.semanticweb.org/maysaa/ontologies/2012/11/untitled-ontology-7#سمع"/>
</owl:ObjectProperty>
```

Figure 4-5 RDFs in Jena tools

## Ontology Examples:

We have two examples first in English word "Mouse", last in Arabic word "Ain" عين .

## 1) Mouse :

Mouse words, in English language contain many concepts. we extracted some meanings from BabelNet show in table 4-1 [61].

Table 4-1. Mouse concept in BabelNet

| Meaning: mouse   •   ID: bn:00056119n   •   Type: Concept | |
|---|---|
| Senses: |  **mouse** <br>  الماوس, |
| Glosses: |  A mouse is a small mammal belonging to the order of rodents, characteristically having a pointed snout, small rounded ears, and a long naked or almost hairless tail. |

| | الفأر جنس من الثدييات تابع لرتبة القوارض.  |
|---|---|
| **Meaning:** shiner • **ID:** bn:00010892n • **Type:** Concept ||
| **Senses:** |  shiner, black eye, **mouse** <br> العين السوداء,  |
| **Glosses:** |  A black eye, periorbital hematoma or shiner, is bruising around the eye commonly due to an injury to the face rather than eye injury. |
| **Meaning:** mouse • **ID:** bn:00056120n • **Type:** Concept ||
| **Senses:** |  **mouse** |
| **Glosses:** |  person who is quiet or timid |
| **Meaning:** mouse • **ID:** bn:00021487n • **Type:** Concept ||
| **Senses:** |  **mouse**, computer mouse <br>  **Mouse (computing)** <br> فأرة  <br> فأرة الكومبيوتر  |
| **Glosses:** |  In computing, a mouse is a pointing device that functions by detecting two-dimensional motion relative to its supporting surface. <br> الفأرة هي إحدى وحدات الإدخال في الحاسوب يتم استعمالها يدويا للتأشير والنقر في  الواجهة الرسومية، وتعتمد أساسا في استعمالها على حركتها فوق سطح مساعد. |
| **Meaning:** Mouse • **ID:** bn:00277032n • **Type:** Concept ||
| **Senses:** |  **Mouse (Alice's Adventures in Wonderland)** <br> الماوس  |
| **Glosses:** |  The Mouse is a fictional character in Alice's Adventures in Wonderland by Lewis Carroll. |

For example, we have two query the *q1* "*Mouse and keyboard*" and *q2* "*Mouse eat corn*s". The proposed model extracted reference concept of queries from the ontologies based on this scenario:

### q1 "*Mouse and keyboard*":

- First, the proposed model will token query to three terms (Mouse term, and term, and keyboard term).

- Next, the stop words will removed like (and) in query.

- Next, the proposed model will process query-processing phase like (stemming).

- Final the proposed model will calculate, the hop counters of classes and properties and relation between two terms (keyboard, mouse). In this query the proposed model, show many classes between terms like computer, farmer, manger etc.

### q2 "*Mouse eat corns*":

- First, the proposed model will token query to three terms (Mouse term, eat term, and corns term).

- Next, the proposed model will process query-processing phase like (stemming). The corns term will change to corn.

- Final the proposed model will calculate, the hop counters of classes and properties and relation between three terms (eat, keyboard, mouse). In this query the proposed model, show many classes between terms like agriculture, computer, farmer, manger etc.

Therefore, the proposed model used shorter root between terms. It catch middle relation between terms. In query, "mouse and keyboard" close to computer more than other classes, properties, and instance in ontology. Thus, the "*computer*" class is reference concept of "mouse and keyboard" query. On the other hand, agriculture class is middle between terms in query 2. Thus, reference concept of q2 is agriculture. In general, the proposed approach able to discrimination reference concept in paragraph. Phrases in paragraph use same methodology that used in extracted RC in queries.

2) **"Ain"** عين :

As shown in table 3 below, one word as " عين Ain (eye) " has a lot of meaning and concepts.

Table 4-2. Eye "Ain" concept

| Glosses | Concept |
|---|---|
| العين هي شبكة كروية وقطر عين الإنسان. <br> Eye, oculus, optic | عضو <br> optic |
| العين هو الحرف الثامن عشر من الألفبائية العربية. <br> Ayin alphabet | حرف ابجدي <br> alphabet |
| العين هي مدينة توجد في أبوظبي. <br> El-Ain city | مدينة <br> City |
| العين الرياضي الثقافي نادٍ رياضي إماراتي. <br> AlAin FC | رياضة <br> Sport |
| عين قانا هي إحدى القرى اللبنانية <br> ElAin village in Lebanon | مدينة <br> City |
| عين مأخوذة من الحسد <br> Envy | سلوك <br> Insanity |
| عين هي التي ينبع منها الماء <br> Appointed | شرب <br> Hole |

For example, we have query " مدينة العين " *Al-Ain city* . The proposed model extracted reference concept of query from the ontologies relies in this scenario:

- First, the proposed model will token query to three terms ( مدينة *city* term, and العين *Al-Ain* term).

- Next, the proposed model will process query-processing phase like (stemming). The العين *Al-Ain* term will change to " عين *Ain* " and " مدينة *Madina*" to " مدن *modn*"

- Final the proposed model will calculate, the hop counters of classes and properties and relation between terms. In this query, proposed model, show many classes between terms like city, hole, sport, etc.

Therefore, the proposed model used shorter root between terms. It catch middle relation between terms. In this query, " مدينة العين " *Al-Ain city* close to geography more than other classes, properties, and instance in ontology. Thus, the "*geography*" class is reference concept of " مدينة العين " *Al-Ain city* query.

## 4.2  Semantic Arabic VSM

In this phase, we implemented two models, traditional model that explained in chapter 2 and the proposed model with semantic. Semantic Arabic VSM relies traditional model and addition reference concept.

VSM or term vector model is an algebraic model for representing text documents and any objects, in general as vectors of identifiers, such as index terms. Figure 4-6 show crawling, indexing and relevancy rankings. In Vector space model documents and queries are represented as vectors.

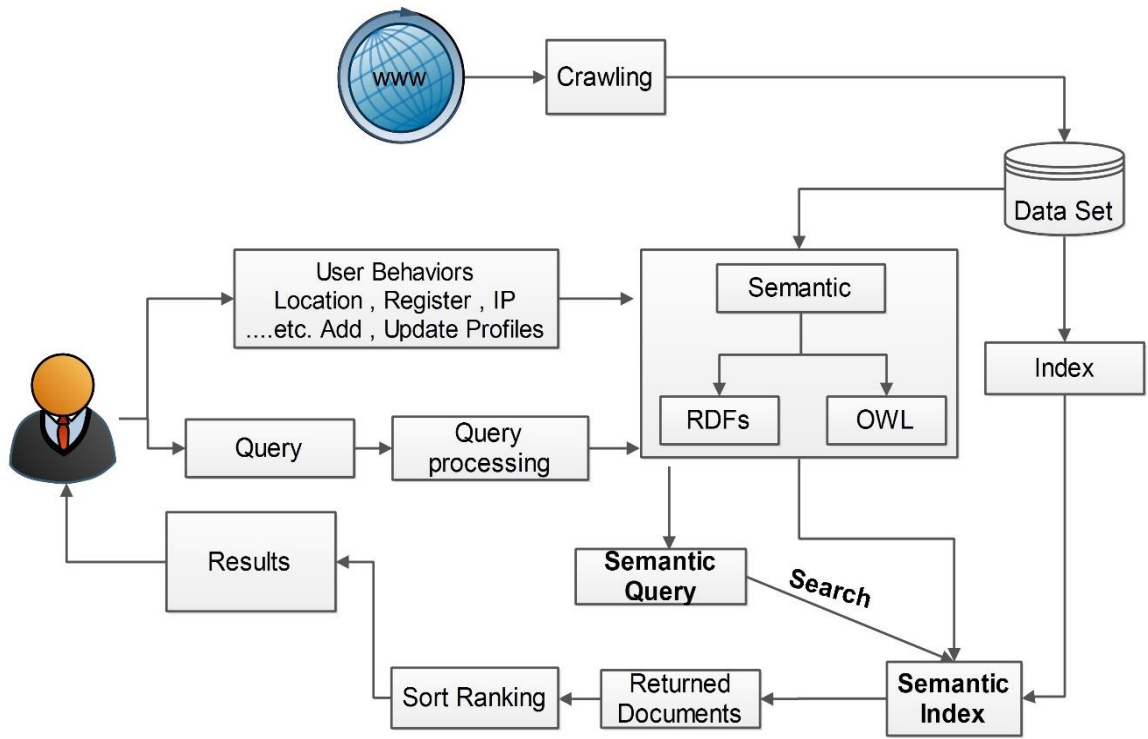

Figure 4-6 Information Retrieval with Semantic model

Each dimension corresponds to a separate term. If a term occurs in the document, its value in the vector is non-zero. Several different ways of computing these values, also known as (term) weights, have been developed. One of the best known schemes is Term Frequency–Inverse Document Frequency *tf-idf* weighting.

Term frequency–inverse document frequency, is a numerical statistic which reflects how important a word is to a document in a collection or corpus [62]. It is often used as a weighting factor in information retrieval and text mining, and it's value increases proportionally to the number of times a word appears in the document, but is offset by the frequency of the word in the corpus, which helps to control for the fact that some words are generally more common than others.

The definition of term depends on the application. Typically terms are single words, keywords, or longer phrases. If the words are chosen to be the terms, the dimensionality of the vector is the number of words in the vocabulary (the number of distinct words occurring in the corpus).

Relevance rankings of documents in a keyword search can be calculated, using the assumptions of document similarities theory, by comparing the deviation of angles between each document vector. The original query vector where the query is represented as the same kind of vector as the documents. To assign a numeric score to a document for a query, the model measures the similarity between the query vector and the document vector.

The similarity between two vectors is once again not inherent in the model. Typically, the angle between two vectors is used as a measure of divergence between the vectors, and cosine of the angle is used as the numeric similarity (since cosine has the nice property that it is 1.0 for identical vectors and 0.0 for orthogonal vectors). Cosine is a measure of similarity between two vectors of an inner product space that measures the cosine of the angle between them [63]. The *tf-idf* weighting is the most common term weighting approach for VSM retrieval is:

$$wtf\text{-}idf_{t,d} = wtf_{t,d} \, . \, idf_t \qquad \textbf{(Equation 4-1)}$$

$$wtf_{t,d} = \begin{cases} 1 + \log_{10} tf_{t,d} & \text{if } tf_{t,d} > 0 \\ 0 & \text{otherwise} \end{cases} \qquad \textbf{(Equation 4-2)}$$

$$idf = \log_{10}(N \, / \, \mathrm{df}_t) \qquad \textbf{(Equation 4-3)}$$

by substitution (2,3) in (1) the weight *tf.idf* is :

$$w_{t,d} = (1 + \log_{10} tf_{t,d}) \times \log_{10}(N \, / \, df_t) \qquad \textbf{(Equation 4-4)}$$

Where $tf_{t,d}$ term frequency of term *t* in document *d* is defined as the number of times, that *t* occurs in *d*. The $df_t$ is the document frequency of *t*: the number of documents that contain t. and $df_t$ is an inverse measure of the in formativeness of *t*, and $df_t \leq$ N, and *idf* Inverse Document Frequency. For the calculating Vector Space and Document Similarity, we have some approaches in similarity measure. One of similarity approaches in equation 4-5 called cosine measure [12]. It is one algorithms to calculate similarity between two documents:

- Each indexing term is a dimension. A indexing term is normally a word.

- Each document is a vector
    - Di = ($t_{i1}$, $t_{i2}$, $t_{i3}$, $t_{i4}$, ..., $t_{in}$)
    - Dj = ($t_{j1}$, $t_{j2}$, $t_{j3}$, $t_{j4}$, ..., $t_{jn}$)

- Document similarity is defined as cosine similarity ($SIM_C$)

$$\text{Similarity } (D_i, D_j) = \frac{\sum_{k=1}^{n} t_{ik} * t_{jk}}{\sqrt{\sum_{k=1}^{n} t_{ik}^2} \times \sqrt{\sum_{k=1}^{n} t_{jk}^2}} \qquad \textbf{Equation 4-5}$$

# Chapter 5
# Experiment Results

# Chapter 5  Experiment Results

This chapter, we have implemented a prototype of the traditional IR model (Boolean and VSM) and the semantic IR model (Boolean and VSM). We have used three main measures to compare between the two models and evaluate our mode. Finally, we have checked the ranking phase in the VSM of the semantic IR model in order to be evaluated correctly. This chapter discusses and explains the Boolean information retrieval model (Section 5.1) and the vector space model (Section 5.2). The Arabic and English language in Boolean model are discussed in Section 5.1.1 and Section 5.1.2 respectively. Finally, the Arabic and English semantic VSM is discussed in Section 5.2.1 and 5.2.2 respectively.

## 5.1  Boolean Information Retrieval Model

In this model we have used a data collection of both Arabic and English language in the implementation of the traditional and semantic IR models. We use the Boolean operators as AND, OR and NOT.

### 5.1.1 English Boolean IR

The proposed model (semantic Boolean  IR) is implemented using Apache Jena which is a Java framework for building semantic Web applications [64].

The obtained results are compared with Lucene which is a high-performance, full-featured text search engine library written entirely in Java [65]. The specification of the platform is Intel core2 Duo 2.10 GHz processor and RAM 3 GB on windows 8.

We used a sample of syntactic dataset. For the sake of testing, samples of two different ontologies are created (device and natural) using Protégé 3.4.3 software [66]. These ontologies will be used in the creation of the semantic index and the searching process as explained in the proposed technique. The precision of the IR model measures the relevant returned documents from all the returned documents and the recall measure the relevant returned documents from the all relevant documents in the collection. Therefore, the lake of semantic in IR models affects only on the precision but the recall will not be affected. Thus, the precision of the proposed semantic IR model and the traditional IR model is measured using Boolean queries with the two Boolean operators (AND, OR).

The results in tables [5-1,5-2] show the precision of the two IR models by using different queries with OR, AND operators respectively. In all previous tested queries, the precision of semantic IR model is always 100%. This is because each word in our dataset has only one ontology concept, which enables the model to detect semantically the required terms. In this model, the precision can be decreased in the case where the word can have more than one ontology concept in its phrase.

Therefore, we can overcome this problem by comparing the two or more ontology concepts of the word with the ontology concepts of the surrounding phrases. In the other side, the average precisions of the traditional IR model with queries of OR an AND operators are 51%, and 54% respectively.

Table 5-1. Precision of traditional IR and semantic
Boolean IR with OR operator queries

| OR | Traditional Model Precision | Semantic Model Precision |
|---|---|---|
| Keyboard or mouse | 25% | 100% |
| Mouse or dog | 80% | 100% |
| Computers or mouse | 50% | 100% |
| Mouse corn | 50% | 100% |
| Average | 51% | 100% |

Table 5-2. Precision of traditional IR and semantic
Boolean IR with AND operator queries

| AND | Traditional Model Precision | Semantic Model Precision |
|---|---|---|
| Keyboard mouse | 50% | 100% |
| Mouse dog | 50% | 100% |
| Computers mouse | 50% | 100% |
| Mouse corn | 67% | 100% |
| Average | 54% | 100% |

The high precision of the semantic IR model is costly in terms of time. The semantic index construction time and the search time are highly increased. This increment is due to the search on ontology to determine the reference ontology concept for each term. Table 5-3 shows the time of traditional IR and semantic Boolean IR with OR operator queries.

Table 5-3. Time of traditional IR and semantic Boolean IR with OR operator queries

| Query | Traditional IR Time (Milliseconds) | Semantic IR Time (Milliseconds) |
|---|---|---|
| Keyboard or mouse | 2 | 212 |
| Mouse or dog | 2 | 137 |
| Computers or mouse | 2 | 158 |
| Mouse or corn | 2 | 198 |
| **Average** | 2 | 176 |

Large in the time consumed in each case is very clear. Therefore, this problem can be solved by using powerful computers which is already exist. In addition, optimization techniques should be developed to decrease the search time in case of semantic Boolean IR models.

5.1.2 **Arabic Boolean IR**

The proposed model (Semantic Boolean Arabic IR) is implemented using Apache Jena which is a Java framework for building Semantic Web applications [64]. The obtained results are compared with Lucene which is a high-performance, full-featured text search engine library written entirely in Java [65]. The specification of the platform is Intel core2 Duo 2.10 GHz processor and RAM 3 GB on windows 8. We used a sample of Arabic syntactic dataset [67]. For the sake of testing, samples of three different Arabic ontologies are created (علوم - إلكترونيات- طبيعة) using Protégé 3.4.3 software [68]. Thus, the precision of the proposed semantic IR model and the traditional IR model is measured using Boolean queries with the three Boolean operators (AND, OR, NOT). The results in tables [5-4,5-5,5-6] show the precision of the two IR models by using different queries with OR, AND, NOT operators respectively. In all cases, the precision of semantic IR model is always 100%. This is because the model can detect semantically the required terms and as a result does not return false results. In the other side, the average precisions the traditional IR model with queries of OR, AND, and NOT operators are 43%, 79%, and 44% respectively.

Table 5-4. Precision of traditional IR and semantic Boolean IR
with OR operator queries

| Queries | Precision | |
|---|---|---|
| | Traditional | semantic |
| أبل أو تفاحة | 25% | 100% |
| مانجو أو تفاحة | 50% | 100% |
| خوخ أو تفاح | 25% | 100% |
| العــــين أو الفراهـــيدي | 33% | 100% |
| العين أو ألم | 67% | 100% |
| السويس أو قناة | 50% | 100% |
| قناة أو المستقبل | 50% | 100% |
| Average | 43% | 100% |

Table 5-5. Precision of traditional IR and semantic Boolean IR with AND operator queries

| Queries | Precision | |
|---|---|---|
| | Traditional | semantic |
| تفاحة و أبل | 50% | 100% |
| تفاحة و بيضاء | 50% | 100% |
| تفاحو مانجو | 100% | 100% |
| العين و للفراهيدي | 100% | 100% |
| ألم و العين | 50% | 100% |
| قناة و السويس | 100% | 100% |
| قناة و المستقبل | 100% | 100% |
| Average | 79% | 100% |

Table 5-6. Precision of traditional IR and semantic Boolean IR with NOT operator queries

| Not | Precision | |
|---|---|---|
| | Traditional | Semantic |
| تفاحة أبل بيضاء ليس خضراء | 33% | 100% |
| العين ليس الفراهيدي | 50% | 100% |
| تفاح ليس قناة | 25% | 100% |
| كتاب ليس العين | 33% | 100% |
| Average | 44% | 100% |

Table 5-7 show the time of traditional IR and semantic Boolean IR with OR operator queries.

Table 5-7. Time of traditional IR and semantic

Boolean IR with OR operator queries

| Query | Traditional IR Time (Milliseconds) | semantic IR Time (Milliseconds) |
|---|---|---|
| أبل OR تفاحة | 3 | 217 |
| مانجو OR تفاحة | 3 | 222 |
| اخضر OR تفاح | 2 | 198 |
| الفراهيديOR العين | 2 | 137 |
| العين OR ألم | 2 | 282 |
| السويس OR قناة | 2 | 114 |
| المستقبلORقناة | 2 | 182 |
| **Average** | 2 | 193 |

Large time consumed in each case is very clear. Therefore, this problem can be solve by using powerful computers which is already exist and in addition, optimization techniques should be developed to decrease the search time in case of semantic Boolean IR models.

## 5.2 Vector Space Model

The semantic VSM model is implemented using Apache Jena which is a Java framework for building semantic Web applications [64]. The obtained results are compared with the traditional VSM model [65].

### 5.2.1 Arabic Vector Space Model

We will processes Arabic queries based on Arabic collection with the two models: Traditional Arabic vector space model (AVSM) and semantic AVSM, thus we discuss ranking results below. For examples, we used three queries:

1. q1 : "تفاحة بيضاء"
2. q2 : "كتاب العين للفراهيدي"
3. q3 : "ألم العين"

#### 5.2.1.1 Traditional VSM model results

We calculate the *tf* for queries from index in tables 5-8,5-9,5-10, by equation 4-2. Next, we calculate *df* and *idf* with in VSM traditional model by equation 4-3. Then, we calculate the *tf-idf* vector for the query by equation 4-4. Final we compute the score of each document in C relative to queries, using the cosine similarity measure by equation 4-5. When computing the *tf-idf* values for the query terms we divide the frequency by the maximum frequency (1) and multiply with the *idf* values.

Table 5-8. *Tf, df* and *idf* in traditional index

for q1"تفاحة بيضاء " terms

| Term | *(docID,tf)* | *df* | *Idf* |
|------|------|------|------|
| بيضاء | (1,1)(3,1)(8,2)(9,1) | 4 | 0.43 |
| تفاح | (1,2)(2,2)(3,10)(6,3)(8,2)(9,1(10,1) | 7 | 0.19 |

Table 5-9. *Tf, df* and *idf* in traditional index for

q2 "كتاب العين للفراهيدي" terms

| Term | *(docID,tf)* | *df* | *Idf* |
|------|------|------|------|
| عين | (1,1)(2,1)(5,16)(6,1)(71)(92(10,6)(11,1) | 8 | 0.13 |
| فراهيدي | (7,4)(10,2)(11,1) | 3 | 0.56 |
| كتاب | (7,1)(9,2)(10,2)(11,2) | 4 | 0.43 |

Table 5-10. *Tf, df* and *idf* in traditional index for

q3 "ألم العين" terms

| Term | *(docID,tf)* | *df* | *Idf* |
|------|------|------|------|
| الم | (2,1)(5,2) | 2 | 0.74 |
| عين | (1,1)(2,1)(5,16)(6,1)(71)(92(10,6)(11,1) | 8 | 0.13 |

We calculate each VSM traditional process (*tf weight* , *tf.idf*, *cosine similarity*) in tables (5-11,5-12,5-13) for queries.

---

Table 5-11. *Wtf*, *tf.idf* and cosine similarity in traditional model
for q1 "تفاحة بيضاء"

| Term | | *ID* 1 | *ID* 2 | *ID* 5 | *ID* 6 | *ID* 7 | *ID* 9 | *ID* 10 |
|---|---|---|---|---|---|---|---|---|
| تفاح | *wtf* | 1.301 | 1.301 | 2 | 1.4771 | 1.3010 | 1 | 1 |
| | *tf.idf* | 0.255 | 0.255 | 0.392 | 0.289 | 0.2553 | 0.196 | 0.196 |
| بيضاء | *wtf* | 1 | 0 | 1 | 0 | 1.301 | 1 | 0 |
| | *tf.idf* | 0.43933 | 0 | 0.43933 | 0 | 0.57158 | 0.439 | 0 |
| $\|d\|=(\sum (tf.idf)^2)^{\frac{1}{2}}$ | | 0.508 | 0.255 | 0.589 | 0.290 | 0.6260 | 0.481 | 0.196 |
| *d.q* | | 0.723 | 0.266 | 0.866 | 0.302 | 0.861 | 0.662 | 0.204 |
| $\|d\|.\|q\|$ | | 0.748 | 0.376 | 0.868 | 0.427 | 0.922 | 0.709 | 0.289 |
| **SIM$_C$** | | **0.967** | **0.707** | **0.998** | **0.707** | **0.934** | **0.934** | **0.707** |

Table 5-12. *Wtf*, *tf.idf* and cosine similarity in traditional model

"كتاب العين للفراهيدي" for q2

| Term | | ID 1 | ID 2 | ID 5 | ID 6 | ID 7 | ID 9 | ID 10 | ID 11 |
|------|------|------|------|------|------|------|------|-------|-------|
| كتاب | *wtf* | 0 | 0 | 0 | 0 | 1 | 1.301 | 1.301 | 1.301 |
| | *tf.idf* | 0 | 0 | 0 | 0 | 0.439 | 0.571 | 0.571 | 0.571 |
| عين | *wtf* | 1 | 1 | 2.204 | 1 | 1 | 1.301 | 1.778 | 1 |
| | *tf.idf* | 0.138 | 0.138 | 0.304 | 0.138 | 0.138 | 0.179 | 0.245 | 0.138 |
| فراهيدي | *wtf* | 0 | 0 | 0 | 0 | 1.602 | 0 | 1.301 | 1 |
| | *tf.idf* | 0 | 0 | 0 | 0 | 0.903 | 0 | 0.734 | 0.564 |
| $\|d\|=(\sum (tf.idf)^2)^{\frac{1}{2}}$ | | 0.138 | 0.138 | 0.305 | 0.138 | 1.015 | 0.599 | 0.962 | 0.815 |
| **d.q** | | 0.144 | 0.144 | 0.317 | 0.144 | 1.543 | 0.783 | 1.616 | 1.327 |
| $\|d\|.\|q\|$ | | 0.249 | 0.249 | 0.550 | 0.249 | 1.830 | 1.081 | 1.736 | 1.470 |
| **SIM$_C$** | | **0.577** | **0.577** | **0.577** | **0.577** | **0.843** | **0.724** | **0.931** | **0.903** |

Table 5-13. *wtf*, *tf.idf* and cosine similarity
in traditional model for q3 " *ألم العين*"

| Term | | ID 1 | ID 2 | ID 5 | ID 6 | ID 7 | ID 9 | ID 10 | ID 11 |
|---|---|---|---|---|---|---|---|---|---|
| ألم | *wtf* | 0 | 1 | 1.3010 | 0 | 0 | 0 | 0 | 0 |
| | *tf.idf* | 0 | 0.7403 | 0.9632 | 0 | 0 | 0 | 0 | 0 |
| عين | *wtf* | 1 | 1 | 2.204 | 1 | 1 | 1.301 | 1.778 | 1 |
| | *tf.idf* | 0.138 | 0.138 | 0.304 | 0.138 | 0.138 | 0.1799 | 0.245 | 0.138 |
| $\|d\|=(\sum (tf.idf)^2)^{1/2}$ | | 0.138 | 0.753 | 1.010 | 0.138 | 0.138 | 0.180 | 0.246 | 0.138 |
| *d.q* | | 0.144 | 0.915 | 1.321 | 0.144 | 0.144 | 0.187 | 0.256 | 0.144 |
| $\|d\|.\|q\|$ | | 0.204 | 1.109 | 1.488 | 0.204 | 0.204 | 0.265 | 0.362 | 0.204 |
| **SIM$_C$** | | **0.707** | **0.825** | **0.888** | **0.707** | **0.707** | **0.707** | **0.707** | **0.707** |

**5.2.1.2 Semantic VSM model results**

We calculate the *tf* for queries terms from semantic index, and we used *df* and *idf* within VSM semantic model in tables 5-14,5-15,5-16 by equations (4-1:4-5).

Table 5-14. *Tf*, *df* and *idf* in semantic model
for q1" تفاحة بيضاء " terms

| Term | RC | (DocId,tf) | Df |
|------|------|------|------|
| بيضاء | لون | (1,1)(3,1)(8,2)(9,1) | 4 |
| تفاح | شعار | (1,2)(8,2) | 2 |
| تفاح | فاكهة | (2,2)(3,10)(6,3)(9,1)10,1) | 5 |

Table 5-15. *Tf*, *df* and *idf* in semantic model
for q2 " كتاب العين للفراهيدي " terms

| *Term* | *RC* | (DocId,tf) | *df* |
|------|------|------|------|
| عين | عضو | (1,1)(2,1)(5,16)(6,1) | 4 |
| عين | حرف | (7,1)(10,6)(11,1) | 3 |
| عين | مدن | (9,2) | 1 |
| فراهيدي | عالم | (7,4)(10,2)(11,1) | 3 |
| كتاب | كتاب | (7,1)(9,2)(10,2)11,2) | 4 |

Table 5-16. *Tf*, *df* and *idf* in semantic model for q3 " ألم العين " terms

| Term | RC | (DocId,tf) | df |
|------|-----|-----------|-----|
| الم | احساس | (2,1)(5,2) | 2 |
| عين | عضو | (1,1)(2,1)(5,16)(6,1) | 4 |
| عين | حرف | (7,1)(10,6)(11,1) | 3 |
| عين | مدن | (9,2) | 1 |

We calculate each VSM semantic process (*tf* weight , *tf.idf*, *cosine similarity*) in tables 5-17,5-18,5-19 for queries.

Table 5-17. *Wtf*, *tf.idf* and cosine similarity in semantic model for q1

| Term | | docID 1 | docID 3 | docID 8 | docID 9 |
|------|-----|---------|---------|---------|---------|
| تفاح | *wtf* | 1.3010 | 0 | 1.3010 | 0 |
| | *tf.idf* | 0.963234 | 0 | 0.9632 | 0 |
| بيضاء | *wtf* | 1 | 1 | 1.301 | 1 |
| | *tf.idf* | 0.439333 | 0.43933 | 0.5716 | 0.43933 |
| $\|d\|=(\sum (tf.idf)^2)^{\frac{1}{2}}$ | | 1.058694 | 0.43933 | 1.1201 | 0.43933 |
| *d.q* | | 1.460623 | 0.45752 | 1.59835 | 0.45752 |
| $\|d\|.\|q\|$ | | 1.559193 | 0.64703 | 1.64957 | 0.64703 |
| SIM$_C$ | | **0.936781** | **0.70711** | **0.96895** | **0.70711** |

Table 5-18. *wtf*, *tf.idf* and cosine similarity in
semantic model for q2

| Term | | ID 7 | ID 9 | ID 10 | ID 11 |
|---|---|---|---|---|---|
| كتاب | *wtf* | 1 | 1.3010 | 1.3010 | 0 |
| | *tf.idf* | 0.439333 | 0.57159 | 0.5716 | 0 |
| عين | *wtf* | 1 | 0 | 1.778 | 1 |
| | *tf.idf* | 0.564271 | 0 | 1.0034 | 0.56427 |
| فراهيدي | *wtf* | 1.6021 | 0.0000 | 1.3010 | 1 |
| | *tf.idf* | 0.903997 | 0 | 0.7341 | 0.56427 |
| $\|d\|=(\sum (tf.idf)^2)^{\frac{1}{2}}$ | | 1.15266 | 0.57159 | 1.36835 | 0.798 |
| d.q | | 1.986562 | 0.59524 | 2.40466 | 1.17526 |
| $\|d\|.\|q\|$ | | 2.079105 | 1.03099 | 2.46816 | 1.43939 |
| **SIM$_C$** | | **0.955489** | **0.57735** | **0.97427** | **0.8165** |

Table 5-19. *Wtf*, *tf.idf* and cosine similarity in

semantic model for q3

| Term | | *ID* 1 | *ID* 2 | *ID* 5 | *ID* 6 |
|---|---|---|---|---|---|
| ألم | *wtf* | 0 | 1 | 1.3010 | 0 |
| | *tf.idf* | 0 | 0.74036 | 0.9632 | 0 |
| عين | *wtf* | 1 | 1 | 2.230 | 1 |
| | *tf.idf* | 0.439333 | 0.43933 | 0.9799 | 0.43933 |
| $||d||=(\sum (tf.idf)^2)^{\frac{1}{2}}$ | | 0.439333 | 0.8609 | 1.37406 | 0.43933 |
| *d.q* | | 0.457518 | 1.22853 | 2.02358 | 0.45752 |
| $||d||.||q||$ | | 0.647028 | 1.26789 | 2.02365 | 0.64703 |
| $SIM_C$ | | **0.707107** | **0.96895** | **0.99996** | **0.70711** |

### 5.2.1.3 Ranking

Table 5-20 shows DocId 3 has a lot of keyword "تفاح" , but we have question, what is meaning of "تفاح" word in this document? fruit or logo!. To answer this question, we need semantic index contain reference concept RC. when we inters "تفاحة بيضاء" query in ontology, the output of ontology: query RC is "logo" not "fruit".

So if we compare between ranking in semantic or traditional, we observe the docId 3 document achieved 1st place in traditional, while it's achieved 3rd place in semantic because "تفاحة بيضاء" query is logo not fruit.

Table 5-20. The Ranking for q1 in traditional and semantic

| Ranking | تفاحة بيضاء | | | |
| | Semantic | | Traditional | |
| | docID | SIM_C | docID | SIM_C |
| 1st | 8 | 0.96895 | 3 | 0.998 |
| 2nd | 1 | 0.936781 | 1 | 0.967 |
| 3rd | 3 | 0.70711 | 9 | 0.934 |
| 4th | 9 | 0.70711 | 8 | 0.934 |

Table 5-21 shows DocId 11 and docId 7 exchanged their positions in ranking table between traditional and semantic. It contain query words "كتاب العين للفراهيدي", this query have "art" concept when we insert it to ontology.

Table 5-21. The Ranking for q1 in traditional and semantic

| Ranking | كتاب العين للفراهيدي | | | |
|---|---|---|---|---|
| | Semantic | | Traditional | |
| | *docID* | SIM$_C$ | *docID* | SIM$_C$ |
| 1st | 10 | 0.974271 | 10 | 0.930877 |
| 2nd | 7 | 0.955489 | 11 | 0.902611 |
| 3rd | 11 | 0.816497 | 7 | 0.843137 |
| 4th | 9 | 0.57735 | 9 | 0.724071 |

Table 5-22 shows docId 11 has a lot of keyword "العين", but we have one question, what is meaning of "العين" word in document 11? Alphabets letter or eye or place!. To answer this question, we need semantic index contains reference concepts RCs.

When we insert "ألم العين" query in ontology, the output of ontology: query RC is "medicine". So if we compare between ranking in semantic or traditional, we observe the docId 11 achieved 3rd place in traditional, while it's not achieved any place in semantic because "ألم العين" query is medicine and docID has only "العين" alphabet letter not eye.

Table 5-22. The Ranking for q1 in traditional and semantic

| Ranking | ألم العين | | | |
| | Semantic | | Traditional | |
| | *docID* | **SIM$_C$** | *docID* | **SIM$_C$** |
|---|---|---|---|---|
| 1st | 5 | 0.99996 | 5 | 0.888 |
| 2nd | 2 | 0.96895 | 2 | 0.825 |
| 3rd | 6 | 0.70711 | 9 | 0.707 |
| 4th | 1 | 0.70711 | 11 | 0.707 |

We have measured the precision and recall of the proposed semantic VSM model. Table 5-23 below shows average of examples achieved high precision of semantic VSM model is more than traditional VSM. Because on semantic model can detect semantically the required terms.

Result of top is improvement precision from 47% in traditional model to 92% in semantic VSM model also traditional model achieved recall 72% in top4 whilst achieved 100% in semantic VSM.

Table 5-23. Precision and Recall of top four in results

| Queries | Traditional | | Semantic | |
|---|---|---|---|---|
| | Recall | Precision | Recall | precision |
| ألم العين | 50 | 50 | 100 | 100 |
| تفاحة بيضاء | 100 | 50 | 100 | 75 |
| كتاب العين للفراهيدي | 66.7 | 40 | 100 | 100 |
| **Average** | 72% | 47% | 100% | 92% |

Table 5-24 below shows the precision measure for each results improvement from 48% to 92% in average of examples.

Table 5-24 Precision and Recall of each results

| Queries | traditional model | | semantic model | |
|---|---|---|---|---|
| | Recall | precision | Recall | precision |
| ألم العين | 100 | 50 | 100 | 100 |
| تفاحة بيضاء | 100 | 45 | 100 | 75 |
| كتاب العين للفراهيدي | 100 | 50 | 100 | 100 |
| **Average** | 100% | 48% | 100% | 92% |

### 5.2.2 **English Vector Space Model**

We will processes queries with two model: English Traditional VSM and Semantic VSM, and we discuss ranking results. For the sake of testing, we have proposed three queries is applied of the collection index.

Query 1 " *mouse eats corn, apple and date*"

Query 2 " *Computer has mouse, keyboard, monitor and system*"

Query 3 " Metropolitan in *Apple*"

As shown in Table 5-25, the documents are very large in collection. This thesis cannot include all terms calculation. Therefore, we will explain a brief about main terms calculation. Mouse term is example in table 5-25.

Table 5-25. Capture from index, example "*mouse*"

| *Term* | *(docID, tf)* |
|--------|---------------|
| **Mouse** | (3,10)(4,4)(6,1)(7,2)(8,18)(12,2)(13,3)(14,2)(15,20)(16,13)(17,19)(18,18)(24,18)(25,5)(26,2)(27,17)(28,12)(29,13)(30,17)(31,13)(32,1)(33,18)(34,14)(35,8)(36,17)(37,17)(38,17)(39,15)(40,9)(41,15)(42,7)(43,10)(44,13)(45,14)(46,10)(47,15)(48,6)(49,12)(50,14)(51,12)(52,3)(53,12)(54,15)(55,4)(56,16)(57,16)(58,3)(59,11)(60,15)(61,11)(62,15)(63,5)(64,12)(65,4)(74,8)(75,16)(76,3)(77,9)(78,14)(79,5)(80,1)(81,4)(82,18)(83,20)(87,4)(88,7)(90,9)(100,17) |

#### 5.2.2.1 Traditional VSM model results

We calculate the *tf* for queries terms from index, and we calculate *df* and *idf* with in VSM traditional model in tables 5-26,5-27,5-28 by equation (4-1:4-5) in chapter 4.

Table 5-26. *Tf*, *df* and *idf* in traditional index

for q1" *mouse eats corn, apple and date* "terms

| *term* | *df* | *Idf* |
|---|---|---|
| Apple | 92 | 0.036212 |
| Corn | 69 | 0.1611 |
| Date | 80 | 0.0969 |
| Eat | 91 | 0.0409586 |
| Mouse | 68 | 0.16749 |

Table 5-27. *Tf*, *df* and *idf* in traditional index

for q2" *Computer has mouse, keyboard, monitor and system* "terms

| *term* | *Df* | *Idf* |
|---|---|---|
| Computer | 57 | 0.24412 |
| Keyboard | 52 | 0.2839 |
| Monitor | 86 | 0.06550 |
| Mouse | 68 | 0.16749 |
| System | 75 | 0.12493 |

Table 5-28. *Tf*, *df* and *idf* in traditional index

for q3" Metropolitan *in Apple* "terms

| Term | df | idf |
|------|------|---------|
| Apple | 92 | 0.03621 |
| Metropolitan | 93 | 0.03151 |

We calculate each VSM traditional process (*tf* weight, *tf.idf*) in tables 5-29,5-30, 5-31 for all documents. Table 5-29 shows docID 1 calculation of *wtf* and *tf.idf*. Other documents in collection use same techniques that used in docID 1 to calculate *wtf* and *tf.idf* .

Table 5-29. *Wtf*, *tf.idf* in traditional model for docID 1

| Doc ID | Terms | wtf | tf.idf |
|--------|-------|-------|--------|
| docID 1 | Metropolitan | 1.954 | 0.0615 |
| | Monitor | 2.113 | 0.1384 |
| | Apple | 1.301 | 0.0471 |
| | Corn | 1.602 | 0.2581 |

**5.2.2.2** Semantic VSM **model** results

Table 5-30 shows, the terms calculate the *tf* for queries from semantic index using Reference Concept RCs, and we calculate  *df* and *idf*  in VSM semantic model .

Table 5-30. DocID and *tf*, In Semantic Model

| Term | RC | (docID, tf) |
|---|---|---|
| **Mouse** | *Animal* | (7,2)(8,18)(12,2)(24,18)(27,17)(28,12)(36, 17)(37,17)(38,17)(39,15)(40,9)(41,15)(50,1 4)(51,12)(61,11)(62,15)(63,5)(74,8)(75,16) (82,18)(83,20)(87,4)(90,9)(100,17) |
| | *Electronic* | (3,10)(4,4)(6,1)(25,5)(26,2)(42,7)(43,10)(4 4,13)(45,14)(46,10)(47,15)(48,6)(49,12)(76 ,3)(77,9)(78,14)(79,5)(80,1) (81,4) |

The VSM semantic process (*df*, *tf.idf*) in table 5-31 for terms of queries. Table 5-31 shows examples of terms in semantic VSM index. The index in semantic add reference concept revers index in traditional model.

Table 5-31. *Df* and *idf* in semantic model for terms with RCs

| Terms | RC | Df | idf |
|---|---|---|---|
| Apple | Town | 41 | 0.387 |
| Apple | Company | 24 | 0.619 |
| Apple | Fruit | 27 | 0.568 |
| Mouse | Animal | 24 | 0.619 |
| Mouse | Electronic | 39 | 0.721 |
| Mouse | Fictional | 5 | 0.602 |

Table 5-32 shows one of documents has terms, RC, *wtf* and *tf.idf.* All documents in collection will use same techniques that used in docID1.

Table 5-32. *Wtf*, *tf.idf* in semantic model for docID 1

| Doc ID | Terms | *RCs* | *wtf* | *tf.idf* |
|--------|-------|-------|-------|----------|
| docID 1 | Metropolitan | Geography | 1.954 | 0.891 |
|  | Apple | Geography | 1.301 | 0.503 |

**5.2.2.3 Ranking**

Tables 5-33,5-34,5-35, shows top 10 of ranking, and notes the different values between traditional VSM compared semantic VSM.

Query 1 retrieved 100 documents in traditional model, while it retrieved 89 documents in semantic model.

Table 5-33. The Ranking for q1
in traditional model and semantic model

| Ranking | Mouse eats corn, apple and date | | | |
| | Traditional | | Semantic | |
| | docID | $SIM_C$ | docID | $SIM_C$ |
|---|---|---|---|---|
| 1st | 8 | 0.57 | 82 | 0.89 |
| 2nd | 50 | 0.57 | 87 | 0.78 |
| 3rd | 54 | 0.56 | 24 | 0.74 |
| 4th | 83 | 0.56 | 74 | 0.73 |
| 5th | 82 | 0.55 | 37 | 0.73 |
| 6th | 56 | 0.55 | 41 | 0.72 |
| 7th | 45 | 0.55 | 65 | 0.72 |
| 8th | 44 | 0.54 | 58 | 0.70 |
| 9th | 90 | 0.53 | 76 | 0.68 |
| 10th | 53 | 0.53 | 75 | 0.66 |

Query 2 return 77 documents from collection in semantic model revers traditional model return 100.

Table 5-34. The Ranking for q2

in traditional model and semantic model

| Ranking | Computer has mouse, keyboard, monitor and system | | | |
| | Traditional | | Semantic | |
| | docID | $SIM_C$ | docID | $SIM_C$ |
|---|---|---|---|---|
| 1st | 62 | 0.95 | 39 | 0.98 |
| 2nd | 44 | 0.93 | 53 | 0.93 |
| 3rd | 15 | 0.93 | 31 | 0.89 |
| 4th | 13 | 0.89 | 40 | 0.89 |
| 5th | 86 | 0.88 | 52 | 0.89 |
| 6th | 76 | 0.88 | 62 | 0.55 |
| 7th | 61 | 0.87 | 8 | 0.31 |
| 8th | 33 | 0.87 | 41 | 0.20 |
| 9th | 8 | 0.86 | 61 | 0.17 |
| 10th | 77 | 0.85 | 86 | 0.10 |

Query 3 return 69 only in semantic while return 100 in traditional model. The 1st document in semantic model not found in traditional model. The sort lists are changed in semantic model due to reference concept and mechanism of approach, which determine the user needs.

Table 5-35. The Ranking for q3 in
traditional model and semantic model

| Ranking | Metropolitan in *Apple* | | | |
| | Traditional | | Semantic | |
| | docID | $SIM_C$ | docID | $SIM_C$ |
|---|---|---|---|---|
| 1st | 69 | 0.11 | 17 | 0.20 |
| 2nd | 55 | 0.11 | 63 | 0.11 |
| 3rd | 15 | 0.11 | 46 | 0.11 |
| 4th | 27 | 0.11 | 65 | 0.09 |
| 5th | 50 | 0.11 | 45 | 0.05 |
| 6th | 95 | 0.11 | 35 | 0.05 |
| 7th | 85 | 0.11 | 44 | 0.04 |
| 8th | 30 | 0.11 | 79 | 0.04 |
| 9th | 48 | 0.11 | 1 | 0.02 |
| 10th | 70 | 0.11 | 59 | 0.02 |

Tables 5-33,5-34,5-35 Shows different between semantic model compare traditional model. The semantic model retrieved only the user need infers traditional model retrieved a lot of result out of users need and out of reference concept.

Traditional model retrieved 987 result from 1000 document in collections for all queries, while semantic model retrieved only 782 document. We observe in tables of traditional model a lot of document's differ the domain of queries.

The average precisions for top 10 in traditional IR model are achieve 66%, and 87% in semantic model.

# Chapter 6
# Conclusion and Future Work

# Chapter 6  Conclusion and Future Work

## 6.1  Conclusion

In this thesis, a new semantic IR model is proposed, this model is based on the use of ontology to represent the relation and the meaning of each word in the index based on it context.

The results show that the new approach enhanced the precision and make it 100% in all cases. On the contrary, the time consumed in the search in the semantic model is very large in compare to the time consumed in the traditional IR models which is not a big problem nowadays because the existence on powerful computing platform.

In addition, the Semantic Vector Space models are implemented. The results show that the new approach enhances the ranking process and the precision the returned results.

We create automatically detect reference concept RC for query from ontology. Another direction is to develop new NLP and optimization techniques to enhance the performance of the creation of the semantic index.

## 6.2 Future work

In future:

- In the future work, optimization techniques will be developed to decrease the construction time and the search time in the semantic Boolean IR models.

- In addition, a semantic ranking IR model will be studied and new ranking techniques will be proposed.

- Optimization techniques will be developed to enhance choose and calculate path among Ontology.

- Building automatic semantic index based on NLP with semantic techniques.

- Create ontology about slang Language.

# Publications

- Emad Elabd, Eissa M. Alshari and Hatem Abdulkader, "Arabic Boolean Information Retrieval based on semantic," in *International Arabic Journal of Information and Communication Technologies,* Acceptance.

- Emad Elabd, Eissa M. Alshari and Hatem Abdulkader, "Boolean Information Retrieval based on semantic," in *Sixth International conference on intelligent computing and information system*, Faculty of computer & Information sciences, Ain Shams University*,* 87-93, Dec 2013.

- Emad Elabd, Eissa M. Alshari and Hatem Abdulkader, "Vector Space Model based on semantic," in *Sixth International conference on intelligent computing and information system*, Faculty of computer & Information sciences, Ain Shams University*,* pp 94-101, Dec 2013.

# References

جامعة المنوفية
كلية الحاسبات والمعلومات
قسم نظم المعلومات

# إطار عمل لاسترجاع المعلومات العربية بالدلالة

رسالة مقدمة كمتطلب جزئي لمتطلبات جامعة المنوفية
للحصول على درجة الماجستير
في نظم المعلومات

## عيسى محمد محسن الشعري

قسم نظم المعلومات
كلية الحاسبات والمعلومات – جامعة المنوفية

## لجنة الإشراف

### د / حاتم محمد سيد أحمد

رئيس قسم نظم المعلومات - وكيل الكلية للدراسات العليا والبحوث – كلية
الحاسبات والمعلومات – جامعة المنوفية

### د / عماد سعيد عبد العليم العبد

مدرس بقسم نظم المعلومات – كلية الحاسبات والمعلومات – جامعة المنوفية

2014

جامعة المنوفية
كلية الحاسبات والمعلومات
قسم نظم المعلومات

# إطار عمل لاسترجاع المعلومات العربية بالدلالة

رسالة مقدمة كمتطلب جزئي لمتطلبات جامعة المنوفية
للحصول على درجة الماجستير
في نظم المعلومات

## عيسى محمد محسن الشعري

قسم نظم المعلومات
كلية الحاسبات والمعلومات – جامعة المنوفية

## لجنــة الحكم المناقشـة

| أ.د/ رأفت عبدالفتاح الكمار | د/ عربي السيد إبراهيم | د/ حاتم محمد سيد أحمد |
|---|---|---|
| أستاذ هندسة الحاسبات، كلية الهندسة، شبرا، جامعة بنها | أستاذ مساعد بقسم علوم الحاسب، عميد كلية الحاسبات والمعلومات، جامعة المنوفية | رئيس قسم نظم المعلومات، وكيل الكلية للدراسات العليا والبحوث – جامعة المنوفية |

**2014**

# ملخـــــص

إن التزايد المستمر في نشر وتخزين المعلومات أظهر صعوبة في عملية استردادها واسترجاعها، وهذا ما أوجد حاجة ملحة إلى تقنيات واطر استرجاع المعلومات بشكل سريع ودقة عالية، إن نظم استرجاع المعلومات حالياً تمتلك الكثير من الطرق لاسترجاع المعلومات وتعد طريقة البحث باستخدام الكلمات المفتاحية keywords هي الطريقة الأكثر انتشاراً في استرجاع المعلومات حالياً، ومع التسارع الهائل في نشر المعلومات ونتيجة للمشاكل اللغوية على سبيل المثال المرادفات وتعدد المعاني في اللغات الحية، جعل من فهم الكلمة ودلالاتها أهمية كبرى في عملية البحث، حيث ان هناك الكثير من الكلمات تحمل معاني ومدلولات كثيرة لا يعرفها ويفرق بينها إلا المستخدم ، مما نتج عنه قصوراً في أنظمة استرجاع المعلومات.

هذه الرسالة تقدم مقترحاً لاسترجاع المعلومات وفهم الدلالات والمعاني يعتمد على نموذج استرجاع المعلومات التقليدي (Information Retrieval) بالإضافة إلى تقنيات الويب الدلالي Semantic web ليكون قادراً على فهم المعاني والمترادفات وتحليل دلالات الكلمات قبل استرجاعها ومن ثم ترتيب النتائج ترتيباً يوافق مع تطلعات واحتياجات المستخدمين، وتختلف طرق استرجاع المعلومات باختلاف النموذج المستخدم حيث تتكون أطر استرجاع المعلومات من مجموعة من النماذج:

- **النموذج البولياني** Boolean Information Retrieval: يعتمد في طريقة عمله على العمليات والقواعد المنطقية مثل (و ، أو ، ليس ... وغيرها ) .

- **النموذج الجبري (الفضاء المتجه)** vector space model: يعتمد على الخوارزميات الجبرية في حساب ترددات الكلمات في المستندات وكذا تردداتها في المجموعة ككل وحساب ندرتها وإعطاء اوزان لكل كلمة على حدة، كما انه يستخدم بعض الخوارزميات في البحث عن التطابق بين

المستندات وبين الاستعلامات مثل الخوارزمية الاقليدية في حساب المتجهات وكذلك خوارزمية استخدام جيب تمام الزاوية في حساب التشابه.

بالإضافة إلى مجموعة من النماذج الأخرى كالنموذج الضبابي والنموذج الاحصائي وغيرها من النماذج التي لم تتناولها الرسالة، حيث يعد النموذجان البولياني والجبري الأكثر شيوعاً في نظم استرجاع المعلومات؛ لذا تم استخدامهما في بناء الإطار المقترح لاسترجاع المعلومات العربية بالاعتماد على الويب الدلالي.

تحتوي أطر استرجاع المعلومات التقليدية على ثلاثة مراحل أساسية في عملية استرجاع المعلومات نوجزها في الآتي:

١) **الفهرسة Indexing** : يتم في هذه المرحلة تخزين المصطلحات وترتيبها وفقاً للنموذج استرجاع المعلومات فني النموذج البولياني يتم تخزين المصطلحات وتخزين وجودها في المستندات بالطريقة المنطقية (موجودة أم لا ) بعيداً عن عدد مرات ورودها بعكس النموذج الجبري الذي يتم فيه تخزين عدد مرات ذكر المصطلح وفي أي مستند تم ذكره.

٢) **معالجة الاستعلام (Query processing)** : في هذه المرحلة يتم تقسيم الاستعلام أو الفقرات إلى عدد من المقاطع والكلمات وبحيث يتم تهذيبها كحذف الكلمات التي تتردد بشكل عام في اللغة أمثال ( في ، عن ، على ) والتي لا تؤثر في عملية الاسترجاع، كما يتم تهذيب الكلمات من خلال تحويلها إلى الجذر الرئيس للكلمة وكذا حذف الحروف الزائدة كاللواحق والزوائد الأمامية للكلمات، وذلك لتقليل عدد الكلمات في الفهرس .

٣) **مرحلة التطابق (Matching)** : في هذه المرحلة يتم مطابقة الاستعلام الخاص بالمستخدم بالفهرس الموجود لدى نظم استرجاع المعلومات وحساب درجة تشابهها وترتيب أولوياتها .

كما تقدم هذه الرسالة نظرة عامة عن الويب الدلالي وعن المكونات الأساسية لتقنيات الويب الدلالي والتي منها أهمها إطار وصف الموارد RDF وكذا مخططات وصف الموارد RDFs والتي تعد اللبنات الرئيسة لبناء أنطولوجي Ontology ، هذه الأنطولوجي تقوم على تمثيل شكلي للمعرفة كمجموعة من المفاهيم ضمن مجال، بالإضافة إلى العلاقات بين تلك المفاهيم. تستخدم الأنطولوجيات للقيام بعمليات تفكير حول كينونات داخل ذلك المجال، وكيف يمكن أن تستخدم لوصف تلك المجالات.

بعد كل ما ذكر عن أطر استرجاع المعلومات وعن ضرورة استعانتها بالويب الدلالي، فإن الرسالة تقدم نموذجاً يدمج بين استرجاع المعلومات التقليدي مع تقنيات الويب الدلالي لتكوين إطار عمل لاسترجاع المعلومات العربية دلالياً، يتكون هذا الإطار المقترح من الآتي:

١) الفهرسة الدلالية: ويتم في هذه المرحلة دمج الفهرس التقليدي بمدلولات الكلمات ومعانيها وفقاً لما يتم استخلاصه من الأنطولوجي.

٢) معالجة الاستعلامات: يتم استخدام الدلالة في معالجة الاستعلام بالإضافة إلى قواعد معالجة الاستعلامات التقليدية.

٣) المطابقة الدلالية: يتم مطابقة الاستعلامات بعد تحويلها إلى استعلامات ذو مدلولات باستخدام تقنيات الويب الدلالي مع الفهارس الدلالية التي يتم انشائها.

إضافة إلى ذلك تقوم الرسالة ببناء أنطولوجي باللغتين العربية والإنجليزي لدعم بناء المقترح، وذلك باستخدام بعض الأدوات القادرة على التعامل مع الويب الدلالي.

وتتوزع هذه الرسالة في ستة فصول نوجزها في الآتي:

**الفصل الأول:** مقدمة عن الرسالة وأهدافها واستعراض المشكلة والحلول المقترحة متضمناً الاسهامات العلمية المقترحة.

**الفصل الثاني:** يتناول هذا الفصل نبذة عن أطر استرجاع المعلومات وانواعها وطريقة عملها وأهدافها وماهية الحاجة إليها، كما يتناول مقدمة عامة عن تقنيات الويب الدلالي مع توضيح لكل مرحلة من مراحل بناء الويب الدلالي.

**الفصل الثالث:** يقدم هذا الفصل نبذة مختصرة عن أهم الدراسات السابقة التي تحدثت عن اطر استرجاع المعلومات وعن علاقتها مع الويب الدلالي سواء الدراسات التي تعالج اللغة العربية او التي تعالج اللغة الإنجليزية وكذا اهم الدراسات التي عالجت تقنيات الويب الدلالي ومراحله وطرق بناءه.

**الفصل الرابع:** يبين هذا الفصل الإطار المقترح لاسترجاع المعلومات دلالية وطريقة بناءه وكيفية عمله واهم القواعد المتبعة فيه، مع تقديم خوارزميات لطريقة عمله متضمناً بعض الأمثلة التي توضح طريقة استخدامه ومدى الاستفادة منه.

**الفصل الخامس:** يتحقق هذا الفصل من نتائج الإطار المقترح والنماذج المنسدلة منه، حيث يقوم بتحليل نتائج الأطر التقليدية بمقارنتها مع نتائج الإطار المقترح في كافة النماذج المستخدمة في الرسالة وإيضاح أهم الجوانب المتعلقة بالنتائج بناءً على مقاييس عالمية في حساب دقة وكفاءة استرجاع المعلومات.

**الفصل السادس:** يقدم نتائج الرسالة والمقترحات المستقبلية.

# موجـز

التزايد المتسارع في تخزين وتوثيق الملفات والمستندات العربية في شبكة الانترنت، أوجد حاجه ملحة لعملية استرجاع المعلومات، وتدعم الكثير من مواقع البحث في الانترنت اللغة العربية في عملية البحث عن المعلومات واستردادها. ولكن النتائج المستخرجة ليست بالكفاءة والدقة والسرعة المطلوبة، في الأغلب تكون غير مكتملة. وذلك يعود إلى أن اللغة العربية معقدة بسبب التراكيب النحوية المعقدة وكذلك كثرة المرادفات والمدلولات وتعدد المعاني. لذا فإن الدافع الرئيس لهذا البحث هو تطوير إطار عمل لاسترجاع المعلومات باللغة العربية يتعرف على المفاهيم الدلالية. وهذا ما تم اعتماده في الإطار المقترح الذي يعتمد على كلاً من النموذج المنطقي ونموذج الفضاء المتجه دلالياً. في الأخير تم تقييم هذا الإطار من خلال مجموعة بيانات لقياس الأداء من بناءً على بعض المعايير القياسية المعروفة لتقييم نظم استرجاع المعلومات مثل مقياس الدقة والاسترجاع والوقت المستهلك. وقد تبين تحسناً ملحوظاً لنتائج الدراسة المعتمدة على الإطار المقترح بمقارنتها بالأطر التقليدية.